\documentclass[aps,prd,twocolumn,showpacs,superscriptaddress,nofootinbib,preprintnumbers]{revtex4-2}

\usepackage[pdftex,colorlinks]{hyperref}
\usepackage[dvipsnames]{xcolor}
\definecolor{phthaloblue}{rgb}{0.0, 0.06, 0.54}

\hypersetup{
    colorlinks=MidnightBlue,
    linkcolor=MidnightBlue,
    citecolor=MidnightBlue,
    filecolor=MidnightBlue,
    urlcolor=MidnightBlue,
    }
\usepackage[pdftex]{graphicx}
\usepackage{bm}
\usepackage{slashed}
\usepackage{cancel}
\usepackage{here}
\usepackage{comment,braket}
\usepackage{epsf}
\usepackage{amsmath}
\usepackage{graphics}
\usepackage{amsfonts}
\usepackage{amssymb}
\usepackage{latexsym}
\usepackage{color}
\usepackage{natbib}
\usepackage[normalem]{ulem}
\input{colordvi.tex}
\usepackage{feynmp}
\usepackage[utf8]{inputenc}
\DeclareUnicodeCharacter{2212}{-}
\usepackage{orcidlink}
\newcommand{\gsim}{\raisebox{-0.13cm}{~\shortstack{$>$ \\[-0.07cm]
      $\sim$}}~}
\newcommand{\lsim}{\raisebox{-0.13cm}{~\shortstack{$<$ \\[-0.07cm]
      $\sim$}}~}

\newcommand{\brac}[2]{ \left( \frac{#1}{#2} \right)}

\newcommand{\beq}{\begin{equation}}
\newcommand{\eeq}{\end{equation}}

\newcommand{\be}{\begin{eqnarray}}
\newcommand{\ee}{\end{eqnarray}}

\usepackage{multirow}
\usepackage{array}
\usepackage{makecell}
\newcommand{\Tstrut}{\rule{0pt}{2.6ex}} % Add vertical spacing in tables

\def\feynrules{{\tt FeynRules}}

\def\cosmotransitions{{\tt CosmoTransitions}}

\begin{document}

\preprint{\tt  FERMILAB-CONF-25-0538-T}

%\title{Probing Inelastic Higgs Portal DM through Indirect Detection and Gravitational Waves}

% \title{Gamma-Ray and Gravitational Wave Signals of Inelastic Higgs Portal Dark Matter}

%\title{Novel signal of Inelastic Higgs Portal DM}
%\title{\Large Novel Probes of Inelastic Higgs Portal DM:\\ \large {From Indirect Detection to Gravitational Waves}}

%\title{The Inelastic Higgs Portal}

\title{Gamma-Rays and Gravitational Waves from Inelastic Higgs Portal Dark Matter}

\author{Dan Hooper\,\orcidlink{0000-0001-8837-4127}}
\email{dwhooper@wisc.edu}

\affiliation{Department of Physics, University of Wisconsin, Madison, WI, USA}

\affiliation{Wisconsin IceCube Particle Astrophysics Center,
University of Wisconsin, Madison, WI, USA}

\author{Gordan Krnjaic\,\orcidlink{0000-0001-7420-9577}}
\email{krnjaicg@fnal.gov}
\affiliation{Theoretical Physics Division, Fermi National Accelerator Laboratory, Batavia, IL, USA}
\affiliation{Department of Astronomy \& Astrophysics, University of Chicago, Chicago, IL USA}
\affiliation{Kavli Institute for Cosmological Physics, University of Chicago, Chicago, IL USA}

\author{Duncan Rocha}
\email{drocha@uchicago.edu}
\affiliation{Theoretical Physics Division, Fermi National Accelerator Laboratory, Batavia, IL, USA}
\affiliation{Department of Physics, University of Chicago, Chicago, IL USA}

\author{Subhojit Roy\,\orcidlink{0000-0001-6434-5268}}
\email{sroy@anl.gov}
\affiliation{High-Energy Physics Division, Argonne National Laboratory, Argonne, IL, USA}

\date{\today}

\begin{abstract}
 We explore  a simple and predictive dark matter scenario involving a {\it complex} scalar field, $\phi$, coupled to the Higgs portal with no additional field content. In the UV, the field possesses a global $U(1)$ symmetry which is broken by mass terms and Higgs portal interactions. 
 In the mass basis, the complex field splits into a pair of real scalars with a small mass splitting (in analogy to pseudo-Dirac fermions), such that the Higgs portal acquires both diagonal and off-diagonal terms with respect to these eigenstates. In the parameter space where the off-diagonal interaction predominates, this scenario is safe from direct detection constraints. Moreover, this model provides a viable explanation for the longstanding Galactic Center gamma-ray excess.
 Additionally, this model influences the Higgs potential in a way that could facilitate a strong first-order electroweak phase transition in the early universe, potentially leading to a stochastic gravitational wave background that could fall within the reach of upcoming space-based detectors.

\end{abstract}

\bigskip
\maketitle

\section{Introduction}
\label{intro}
One of the simplest and most predictive dark matter (DM) models involves a singlet scalar field coupled to the Higgs portal~\cite{Silveira:1985rk,McDonald:1993ex,Burgess:2000yq,OConnell:2006rsp,Cline:2013gha,Duerr:2015mva,Duerr:2015aka,Duerr:2015bea,Hooper:2018buz,Fraser:2020dpy},
\be
\label{eq:min-portal}
{\cal L} \supset -\frac{1}{2} \left( m_{\phi}^2 + k H^\dagger H  \right)\phi^2,
\ee
where $\phi$ is a real scalar that is stabilized by a $Z_2$ symmetry, $H$ is the Standard Model Higgs doublet, and $\kappa$ is the strength of the portal coupling. If $k$ is sufficiently large, the $\phi$ population will thermalize with the Standard Model bath in the early universe and then freeze out of chemical equilibrium, leaving behind a thermal relic abundance.

If $m_{\phi} \gg m_h$, DM annihilation can proceed in this model through $\phi \phi \to h \to W^+ W^-$, $\phi \phi \to h \to ZZ$, and $\phi \phi \to hh$, collectively leading to the following thermal relic abundance~\cite{Duerr:2015aka}:
\be
\Omega_\phi h^2 \approx \Omega_{\rm DM} h^2 \, \bigg(\frac{0.15}{k}\bigg)^2 \bigg(\frac{m_\phi}{\rm TeV}\bigg)^2,
\ee
where $\Omega_{\rm DM} h^2  \approx 0.12$ is the measured DM abundance~\cite{Planck:2018vyg}.  
In the opposite regime, $m_{\phi} \lsim m_h$, the dominant annihilation channels proceed through $s$-channel Higgs exchange to Standard Model fermions or gauge bosons. 

For each choice of the DM mass in this scenario, there is a unique value of $k$ that yields the measured DM abundance. This value of $k$, in turn, determines the magnitude of the signals that are predicted at direct and indirect detection experiments, as well as at colliders. With the exception of the region near the Higgs resonance, $m_\phi \approx m_h/2$, where an acceptable relic abundance can be achieved for small values of $k$, this model is ruled out by a combination of direct detection~\cite{Cline_2013, Casas:2017jjg, DiazSaez:2024nrq}, indirect detection~\cite{DeLaTorreLuque:2023fyg}, and collider constraints~\cite{Djouadi:2011aa,Arcadi:2019lka,Krnjaic:2015mbs}. 

In this paper, we investigate a scenario in which the Higgs portal DM candidate, $\phi$, is promoted to a {\it complex} scalar field. If all allowed mass terms are present, the complex field splits into a pair of non-degenerate mass eigenstates, $\phi_{1,2}$.
For some choices of the masses and couplings, the dominant Higgs portal interaction is off-diagonal in the mass basis,
\be
{\cal L} \supset g \phi_1 \phi_2 H^\dagger H\,,
\ee
where $g$ is a dimensionless coupling. This operator leads to {\it inelastic} scattering at direct detection experiments, greatly weakening constraints from nuclear recoil searches.

The inelastic scalar Higgs portal DM scenario %involving two real singlets 
has been previously explored in Refs.~\cite{Ghorbani:2014gka, Casas:2017jjg, Guo:2021vpb, DiazSaez:2024nrq, Casas:2017rww}. In this work, we revisit this model's predictions in light of recent constraints from direct detection experiments. We further examine the ability of this scenario to explain the spectrum, angular distribution, and overall intensity of the observed excess of GeV-scale emission from the Galactic Center~\cite{Fermi-LAT:2017opo,McDaniel:2023bju,DiMauro:2022hue}, as well as the excess of $\sim 10-20 \, \text{GeV}$ antiprotons in the cosmic-ray spectrum~\cite{Cuoco:2016eej,Giesen:2015ufa,Cholis:2019ejx}.

In order for the matter-antimatter asymmetry of our universe to have been generated through electroweak baryogenesis~\cite{Trodden:1998ym, Anderson:1991zb, Huet:1995sh, Morrissey:2012db}, 
there must have been a strong first-order phase transition in the Higgs field. This can be achieved by introducing extra scalar degrees-of-freedom which modify the Higgs potential.  
In this work, we show that the inclusion of the complex scalar field, $\phi$, can significantly impact the shape of the Higgs potential, allowing for a strong first-order phrase transition. Such a phase transition would generate a stochastic background gravitational waves~\cite{Athron:2023xlk} that could potentially be detectable by future space-based gravitational wave observatories.

The remainder of this paper is organized as follows. In Sec.~\ref{model}, we discuss the theoretical framework of the inelastic Higgs portal model. 
In Secs.~\ref{DD} and ~\ref{relic}, we consider the constraints from direct detection experiments and calculate the thermal relic abundance of DM in this model. In Secs.~\ref{phi2decay} and~\ref{ID}, we consider the cosmological constraints on this model and discuss its ability to produce a signal that is consistent with the Galactic Center Gamma-Ray Excess. The electroweak phase transition (EWPT) and the production of gravitational waves in this model are described in Secs.~\ref{subsec:ewpt-gw}-\ref{sec:gravitationalwaves}. Collider probes of the relevant parameter space in this model are discussed in Sec.~\ref{collider}. We summarize our results and present our conclusions in Sec.~\ref{Conclusions}. In a series of appendices, we discuss contributions to the DM's elastic scattering cross section at one-loop level, the Higgs potential at finite temperature, and the production of gravitational waves from a first-order phase transition in the early universe.
\section{Model Overview}
\label{model}
The tree-level Standard Model Higgs potential can be expressed as follows:
\begin{align}
\label{Vsm}
V_{\text{SM}} = -\mu_h^2H^{\dagger}H + \lambda_h (H^{\dagger}H)^2, 
\end{align}
where $H$ is the Standard Model Higgs doublet. In our scenario, we introduce a complex scalar field, $\phi$, with fully general mass terms. Prior to electroweak symmetry breaking, the Lagrangian includes the following terms:
\be
\label{eq:Lm}
{\cal L}_{m} = -m_0^2 |\phi|^2  -\frac{1}{2}( \rho_0^2 \phi^2 + \rho_0^{*2} \phi^{*2} )\,,
 \ee
where $m_0$ is a real mass parameter, $\rho_0$ is a complex dimensionful scale, and there is a $Z_2$ symmetry under which $\phi \to -\phi$. We also include corresponding bilinear Higgs portal couplings,
\be
\label{eq:Lh}
{\cal L}_{\phi h}  =  -\left[    
\kappa |\phi|^2  +  \frac{1}{2} ( \eta \phi^2   + \eta^* \phi^{*2} )
\right] H^\dagger H \, ,
\ee
where $\kappa$ is a real coupling, and $\eta$ is a complex coupling. 
In addition to the mass terms and Higgs portal interactions, the most general renormalizable scalar potential also contains quartic self-interactions of the singlet fields, which will play an important role in determining the early Universe phase transition dynamics. For clarity, we defer the full expression of these terms to Sec.~\ref{subsec:ewpt-gw}, where we analyze the electroweak phase transition in detail.
After electroweak symmetry breaking, the mass terms  from Eq.~\ref{eq:Lm} become
\be
\label{eq:Lm_v}
{\cal L}_{m} \to -m^2 |\phi|^2  -\frac{1}{2}( \rho^2 \phi^2 + \rho^{*2} \phi^{*2} )\,,
 \ee
where we have defined the parameters
\be
m^2 \equiv m_0^2 +\frac{ \kappa v_h^2 }{2}\,, \,\,\,\,\,\,
\rho^2 \equiv \rho_0^2 + \frac{\eta v_h^2}{2} \, ,
\ee
which include contributions from electroweak symmetry breaking, and where $v_h \approx 246$ GeV is the vacuum expectation value (vev) of the Higgs field. This set of mass terms and Higgs portal interactions can arise if $\phi$ enjoys an approximate $U(1)$ global symmetry under which $\phi \to e^{i\theta} \phi$. We assume this symmetry is broken only by the $\phi^2$ terms in the Lagrangian, and prohibit linear $\phi$ interactions that would induce DM decay.\footnote{Avoiding linear interactions can be achieved if, for example, the complex field, $\phi$, has a charge of $+1$ under the $U(1)$ and our desired breaking terms arise from the vev of an additional scalar, $\Phi$, with a charge of $+2$. This allows terms of the form $\Phi \phi^2 \to \langle\Phi \rangle \phi^2$ to arise without generating any terms that are linear in $\phi$ because, when $\Phi$ acquires a vev, there is a residual $Z_2$ symmetry under which the real mass eigenstates transform as $\phi_i \to -\phi_i$.}

After diagonalizing the mass matrix in Eq.~\ref{eq:Lm}, 
the mass eigenstates of the theory are $\phi_{1}$ and $\phi_2$, with corresponding eigenvalues that are given by
\be
m^2_{\phi_1, \phi_2} = m^2 \mp |\rho^2| \, .
\ee
In the limit of a small mass splitting, we have
\be
 \Delta m = m_{\phi_2} - m_{\phi_1} \approx \frac{|\rho^2|}{ m} \, .
\ee 
The fields in the interaction basis can be written as
\be
\phi = \alpha \phi_1 + \beta \phi_2\, , \,\,\,\,\,\,\,\,
\phi^* = \alpha^* \phi_1 + \beta^* \phi_2\,,
\ee
where $\alpha = e^{-i \theta}/\sqrt{2}$, $\beta = i e^{-i \theta}/\sqrt{2}$, and $\theta$ is the angle that diagonalizes the matrix associated with Eq.~\ref{eq:Lm_v}: 
\beq
\label{eqn:mneut}
{\cal U} =
\left( \begin{array}{ccccc}
\cos\theta & - \sin\theta  \\[0.4cm]
\sin\theta & \cos\theta 
\end{array} \right),
\eeq
where
$\cos2\theta = \mathrm{Re}(\rho^2)/|\rho^2|$ and $\sin2\theta = \mathrm{Im}(\rho^2)/|\rho^2|$.

After electroweak symmetry breaking, the Higgs portal interactions given in Eq.~\ref{eq:Lh}, written in the mass basis, become 
\be
\label{eq:Lh-mass}
{\cal L}_{\phi h} = - \biggl(f_1 \phi_1^2 + f_2 \phi_2^2 + g \phi_1 \phi_2  \biggr)\!\!\left(  -v_h h + \frac{h^2}{2} \right)\, , \,\,
\ee 
where we have defined the diagonal (elastic)  
couplings, 
\be
\label{f1rel}
f_1 &=& \frac{\kappa}{2} + \frac{1}{2} ( \eta \alpha^2  + \eta^* \alpha^{*2}  ) \nonumber
\\
\label{f1rel}
&=& \frac{\kappa}{2}  +  \frac{1}{2}\mathrm{Re}(\eta) \cos2\theta +  \frac{1}{2} \mathrm{Im}(\eta) \sin2\theta  \, ,~~~ \\
f_2 &=& \frac{\kappa}{2} + \frac{1}{2} ( \eta \beta^2  + \eta^* \beta^{*2}  ) \nonumber \\
&=& \frac{\kappa}{2} - \frac{1}{2} \mathrm{Re}(\eta) \cos2\theta - \frac{1}{2} \mathrm{Im}(\eta) \sin2\theta \, ,\ee
and the off-diagonal (inelastic) coupling, 
\label{f2rel}
\be
\label{grel}
g  = \eta \alpha \beta + \eta^{\star} \alpha^{\star} \beta^{\star}
=   \mathrm{Re}(\eta) \sin2\theta -  \mathrm{Im}(\eta) \cos2\theta \,. ~~~~~
\ee

In regions of parameter space with $|f_1| \ll |g|, |f_2|$, this model deviates qualitatively from the familiar case of DM in the form of a real scalar that couples to the Standard Model through the Higgs portal. 
\section{Direct Detection}
\label{DD}
In the traditional Higgs portal scenario characterized by Eq.~\ref{eq:min-portal}, the DM is a real scalar which undergoes efficient elastic scattering with nuclei. For values of the portal coupling, $\kappa$, that give rise to the observed relic density, this elastic scattering cross section is excluded by existing direct detection experiments~\cite{LZCollaboration:2024lux}, except in the fine-tuned Higgs resonance window near $m_{\phi} \approx m_h/2$.

By contrast, in our version of the Higgs portal scenario, the DM can scatter elastically ($\phi_1 N \rightarrow \phi_1 N$) through the coupling, $f_1$, or inelastically ($\phi_1 N \rightarrow \phi_2 N$) through the coupling, $g$. The latter process is only possible if the mass splitting between the two states satisfies the inequality~\cite{Tucker-Smith:2001myb},
\begin{align}
\Delta m < \frac{\mu_{\phi_1 N} }{2} v^2,
\end{align}
where $v$ is the velocity of the DM particle and $\mu_{\phi_1 N}$ is the reduced mass of the DM-nucleus system. For the case of a xenon target and $v \sim 300 \, {\rm km/s}$, inelastic scattering can occur only if $\Delta m \lsim  50 \, {\rm keV}$. Throughout this study, we will restrict ourselves to parameter space in which $\Delta m \gsim {\rm MeV}$, for which inelastic scattering will be prohibitively suppressed.

\begin{figure}[t]
    \centering
        \includegraphics[width=0.99\linewidth]{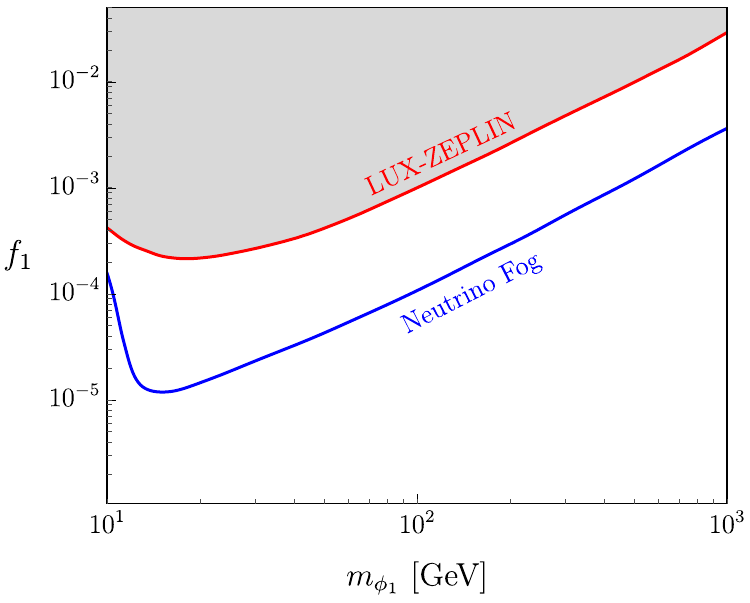}
    \caption{  
    The maximum value of $f_1$ consistent with constraints from the LUX-ZEPLIN (LZ) experiment~\cite{LZCollaboration:2024lux}. Also shown is the value of this parameter that would lead to an elastic scattering cross section at the boundary of the the so-called ``neutrino fog''~\cite{Billard:2013qya}.
    } 
    \label{fig:direct}    
\end{figure}

In this scenario, the spin-independent cross section for elastic scattering off nuclei can be written as 
\begin{align}
\sigma_{\phi_1 N}^{\rm SI} &= \frac{f_1^2 \, m^2_{N} \mu^{2}_{\phi_1 N}}{\pi m_{\phi_1}^2 m^4_h} \Bigg(\sum_{q=u,d,s}f_{T_q}^{N} + \frac{2}{27} \sum_{q=c,b,t} f^{N}_{\rm TG}\Bigg)^2 \nonumber \\
&\approx 4 \times 10^{-46} \, {\rm cm}^2 \, \bigg(\frac{f_1}{0.1}\bigg)^2 \bigg(\frac{\rm TeV}{m_{\phi_1}}\bigg)^2,
\end{align}
where $N =p,n$, $\mu_{\phi_1 N}$ is the $\phi_1$\,-$N$ reduced mass, and the terms in the sums are nuclear matrix elements.
Here, $f^N_{T_q}$ denote the nucleon matrix elements of the light quark scalar operators, defined by $m_N f^N_{T_q} \equiv \langle N | m_q \bar{q} q | N \rangle$, while $f^N_{TG} = 1 - \sum_{q=u,d,s} f^N_{T_q}$ accounts for the gluon contribution via the trace anomaly.~\footnote{From Ref.~\cite{Hill:2014yxa}, we take
$f^p_{T_u} = 0.018$, $f^p_{T_d} = 0.027$, $f^p_{T_s} = 0.037$, $f^n_{T_u} = 0.013$, $f^n_{T_d} = 0.040$, $f^n_{T_s} = 0.037$, and $f_{\rm TG}^{p,n} = 0.91$.
}
The null results of direct detection experiments allow us to place stringent constraints on the coupling, $f_1$, and thus require $|f_1| \ll |g|$ to obtain an acceptable thermal relic abundance. In Fig.~\ref{fig:direct}, we show the maximum value of $|f_1|$ that is consistent with the recent limits placed by the LUX-ZEPLIN (LZ) Collaboration~\cite{LZCollaboration:2024lux}, as a function of $m_{\phi_1}$. This constraint meaningfully impacts the regions of parameter space that are potentially viable within this model. In this analysis, we have used the publicly available package, $\feynrules$~\cite{Alloul:2013bka}, which we interface with~\texttt{MicrOMEGAs}~\cite{Belanger:2006is} to obtain the DM relic abundance, as well as the cross sections relevant for direct and indirect searches.

For completeness, we note that there are one-loop contributions to the $\phi_1 N \to \phi_1 N$ elastic scattering cross section which depend on $g$ and $f_2$, rather than $f_1$. The corresponding Feynman diagrams involve a vertex correction to the $\phi_1 \phi_1 h$ interaction induced by three propagators from virtual $\phi_2$ and $h$ internal lines (see Appendix~\ref{oneloopDD}).
For the benchmark models presented in Table \ref{tab:BPs}, we have verified that these contributions are subdominant.
We also note that direct detection signals could be enhanced in this model if the coefficient of the $\lambda_{12} \phi^2_1\phi_2^2$ quartic operator is large. This coupling primarily impacts the EWPT and the corresponding gravitational wave signal (see Sec.~\ref{subsec:ewpt-gw}), and does not play a significant role in determining the DM relic density.
\section{Thermal Relic Abundance}
\label{relic}
In the early universe, $T \gg m_{\phi_1}, m_{\phi_2}$, the $\phi_{1}$ and $\phi_2$ populations were each maintained in chemical equilibrium with the Standard Model radiation bath. As the universe expanded and cooled, these species froze out of chemical equilibrium, leaving behind a thermal relic abundance. In the case of $\Delta m \ll T_{\rm FO} \approx m_{\phi_1}/20$, the mass splitting between these two states was negligible during the process of thermal freeze out, allowing both annihilations and coannihilations to deplete the abundances of these particles. For larger mass splittings, the thermal relic abundance is largely determined by rate of $\phi_1$ annihilations.

In this scenario, the $\phi_1$ population evolves according to the Boltzmann equation,
\begin{align}
\dot n_{\phi_1} =- 3 H n_{\phi_1}  
&- \langle \sigma v\rangle_{\phi_1\phi_1} \left[ n^2_{\phi_1}  - (n_{\phi_1}^{\rm eq})^2 \right] \\
&- \langle \sigma v\rangle_{\phi_1\phi_2 } \left[ n_{\phi_1} n_{\phi_2} - n_{\phi_1}^{\rm eq} n_{\phi_2}^{\rm eq} \right], \nonumber
\end{align}
and the $\phi_2$ abundance evolves according to
\begin{align}
\dot n_{\phi_2} =- 3 H n_{\phi_2}  
&- \langle \sigma v\rangle_{\phi_2\phi_2 } \left[ n^2_{\phi_2}  - (n_{\phi_2}^{\rm eq})^2 \right] \\
&- \langle \sigma v\rangle_{\phi_1\phi_2} \left[ n_{\phi_1} n_{\phi_2} - n_{\phi_1}^{\rm eq} n_{\phi_2}^{\rm eq} \right], \nonumber
\end{align}
where the ``eq'' superscript denotes an equilibrium quantity, $H$ is the Hubble rate, and the cross sections are thermally averaged and summed over all Standard Model final states. Since processes converting $\phi_1 \leftrightarrow \phi_2$ are efficient during this time, the quantities $n_{\phi_1}/(n_{\phi_1}+n_{\phi_2})$ and $n_{\phi_2}/(n_{\phi_1}+n_{\phi_2})$ remain near their equilibrium values throughout thermal freeze-out, 
allowing these Boltzmann equations to reduce to 
\begin{align}
\dot n_{\rm tot} =- 3 H n_{\rm tot}  
- \langle \sigma v\rangle_{\rm eff} \left[ n^2_{\rm tot}  - (n_{\rm tot}^{\rm eq})^2 \right], 
\end{align}
where $n_{\rm tot} \equiv n_{\phi_1}+n_{\phi_2}$, and the effective annihilation cross section is given by
\begin{align}
\label{eq:effsigma1}
\langle \sigma v \rangle_{\rm eff} &= 
\frac{\langle \sigma v\rangle_{\phi_1\phi_1 }+\epsilon^2 \langle \sigma v\rangle_{\phi_2\phi_2 } \, +2 \epsilon\, \langle \sigma v\rangle_{\phi_1\phi_2 }}{(1+\epsilon)^2}  \, ,
\end{align}
where we have defined the parameter
\be
\epsilon \equiv  \frac{n^{\rm eq}_{\phi_2}}{n^{\rm eq}_{\phi_1}} = 
\left(1+\frac{\Delta m}{m_{\phi_1}} \, \right)^{3/2} e^{- \Delta m/T}.
\ee
Each term in Eq.~\ref{eq:effsigma1} is shorthand for a total
annihilation/coannihilation cross section into all kinematically accessible Standard Model final states, 
\be
\langle \sigma v\rangle_{\phi_i \phi_j} &\equiv&
\langle \sigma v\rangle_{\phi_i \phi_j \to WW}  + 
\langle \sigma v\rangle_{\phi_i \phi_j \to ZZ}  
\nonumber \\
&&+
\langle \sigma v\rangle_{\phi_i \phi_j \to hh}  + \sum_f \langle \sigma v\rangle_{\phi_i \phi_j \to f \bar f} ~.
\ee
In the limit of a small mass splitting, the effective cross section reduces to
\be
\label{eq:effsigma}
\langle \sigma v\rangle_{\rm eff} \approx \frac{1}{2} 
\langle \sigma v\rangle_{\phi_1\phi_2}
+\frac{1}{4}\biggl(
\langle \sigma v\rangle_{\phi_1\phi_1} 
+\langle \sigma v\rangle_{\phi_2\phi_2}
\biggr). \,\,\,\,
\ee

Under the assumption of a negligible mass splitting, $m_{\phi_1}=m_{\phi_2} \equiv m_\phi$, the annihilation and coannihilation cross sections to fermionic final states are given by
\be
\sigma_{\phi_i \phi_i \to f \bar f} &=& 
\frac{n_c f_i^2}{2 \pi s} 
    \sqrt{
    \frac{s -4m_f^2}{s - 4 m_\phi^2}} \left[\frac{m_f^2 \, (s-4m^2_f)}{(s-m_h^2)^2 + m_h^2 \Gamma_h^2}\right]\!\!,~~~~~~  \\
\sigma_{\phi_1 \phi_2 \to f \bar f} &=& \frac{n_c g^2}{8 \pi s} 
    \sqrt{
    \frac{s -4m_f^2}{s - 4 m_\phi^2}} \left[ \frac{m_f^2 \, (s-4m^2_f)}{(s-m_h^2)^2 + m_h^2 \Gamma_h^2} \right]\!,~~~~~~
\\ \nonumber
\ee
where $i = 1,2$ and $n_c = 1 (3)$ is the number of colors of the final state lepton (quark). 
If $m_{\phi} > m_h$, annihilations and coannihilations can also proceed through $t$-channel $\phi$ exchange and through a $\phi \phi h h$ vertex, resulting in the following cross sections, also given in the limit of equal $\phi_{1}$ and $\phi_2$ masses:

\begin{widetext}
\be
    \sigma_{\phi_i \phi_i \to h h} &=& \frac{1}{8 \pi s} 
    \sqrt{
    \frac{s -4m_h^2}{s - 4 m_{\phi}^2}} \left[ f_i^2 w(s)^2 + \frac{2(g^2 + 4f_i^2)^2v_h^4}{A^2 - B^2} + \frac{(g^2 + 4 f_i^2) v_h^2}{AB} [(g^2 + 4 f_i^2) v_h^2 - 2f_i Aw(s)] \ln \bigg(\frac{A+B}{A-B}\bigg) \right]\!,~~~~~ \\
\sigma_{\phi_1 \phi_2 \to h h} &=& \frac{g^2}{32 \pi s} 
    \sqrt{
    \frac{s -4m_h^2}{s - 4m_{\phi}^2}} \left[ w(s)^2 + \frac{32(f_1 + f_2)^2 v_h^4}{A^2 - B^2} + \frac{8(f_1 + f_2) v_h^2}{AB}\left[2 (f_1+f_2)v_h^2 -Aw(s)\right]\ln\bigg(\frac{A+B}{A-B}\bigg) \right], 
\ee
\end{widetext}
where $w(s) = (1 + 2 m_h^2/s) /(1-m_h^2/s)$. Additionally, we have defined 
\be
A \equiv s -2m_h^2 \,,~~~~ 
B \equiv \sqrt{ (s - 4m_{\phi}^2) (s - 4m_h^2)}.~
\ee
Lastly, in the region in which $m_{W/Z} < m_{\phi_i} < m_h$,
annihilations to $W$ and $Z$ bosons can be the dominant reaction.
%, lending to the sizable electroweak couplings. 
In the limit of equal scalar masses, the coannihilation cross sections to a pair of gauge bosons are given by
\be
\sigma_{\phi_1 \phi_2 \to WW} &=& \frac{g^2}{16 \pi s} \sqrt{ \frac{s - 4m_W^2}{s - 4 m_\phi^2} } \frac{8 m_W^4 + \left(s - 2 m_W^2\right)^{\!2} }{(s-m_h^2 )^2 + m_h^2 \Gamma_h^2} \, , \,\,\,\,\,\,\,\,\,\,\,\,\,
\\
\sigma_{\phi_1 \phi_2 \to ZZ}\hspace{5pt} &=& \frac{g^2}{32 \pi s} \sqrt{ \frac{s - 4m_Z^2}{s - 4 m_\phi^2} } \frac{8 m_Z^4 + \left(s - 2 m_Z^2\right)^{\!2}
}{(s-m_h^2 )^2 + m_h^2 \Gamma_h^2} \,.
\ee 
The cross sections for $\phi_i \phi_i \to WW/ZZ$ can be computed by replacing $g \to 2 f_i$ in the above equations.

From the cross sections given above, we can calculate the thermally-averaged value following the standard procedure~\cite{Gondolo:1990dk,Krnjaic:2017tio},
\be
\label{eq:gondolo}
\langle \sigma v \rangle_{\phi_i \phi_j }
=
\frac{1}{N(T)} \!\! \int_{s_0}^\infty \! ds  \, \sigma_{\phi_i \phi_j} \sqrt{s} (s-s_0)   K_2\brac{\sqrt{s}}{T} \!,~~~~~~
\ee
where $s_0 = (m_{\phi_i} + m_{\phi_j} )^2$ is the COM energy at zero momentum, $\sigma_{\phi_i \phi_j}$ is the total annihilation cross section to all SM species, and
\be
N(T) = 8 m^2_{\phi_i} m^2_{\phi_j} T K_1\brac{m_{\phi_i}}{T}  K_1\brac{m_{\phi_j}}{T},
\ee
is a normalization factor.
\begin{figure}[t!]
    \centering
    \includegraphics[width=0.98\linewidth]{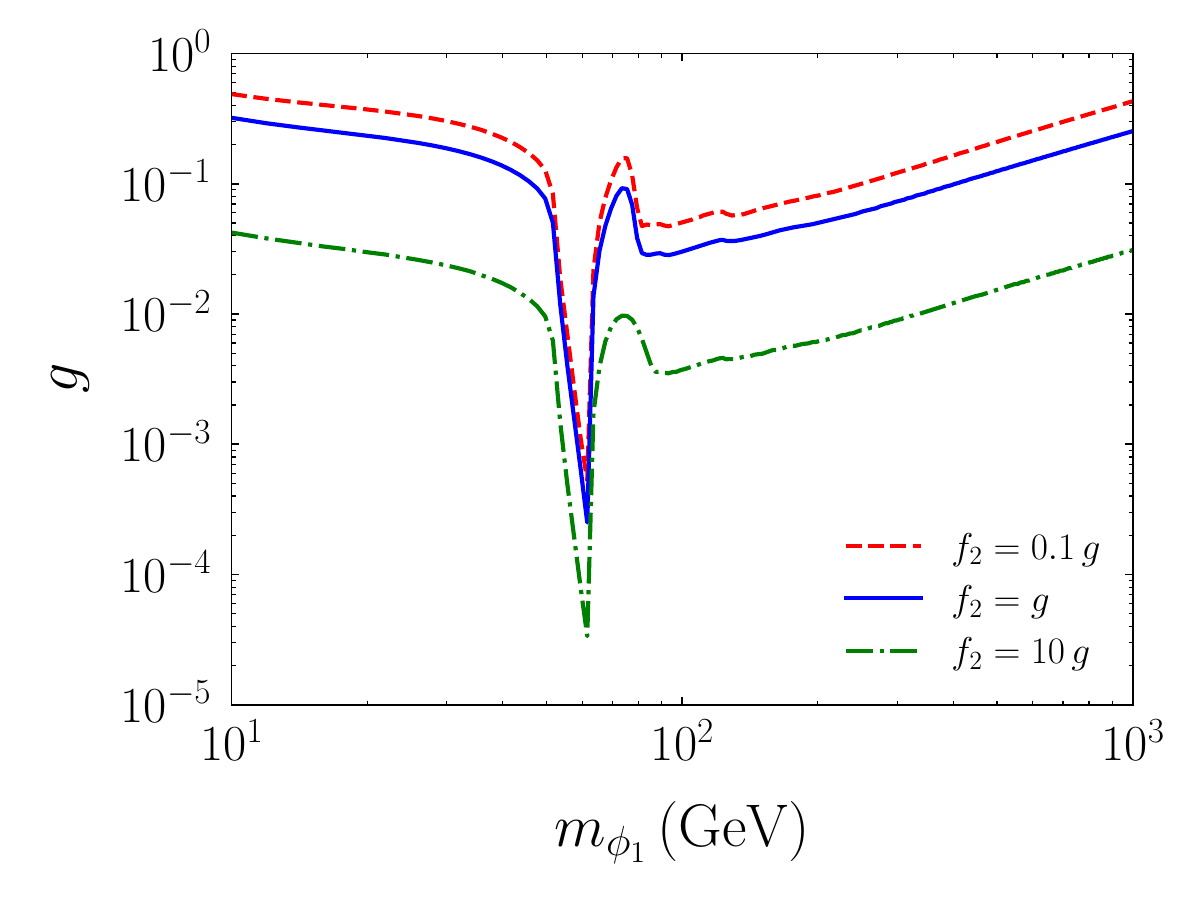}
    \caption{The value of the coupling, $g$, that yields a thermal relic abundance of $\Omega_{\phi_1} h^2 = 0.12$, for $\Delta m = 0.1$~GeV and $f_1 = 0$. Results are shown for three values of $f_2/g$.}
    \label{fig:relic}
\end{figure}

In Fig.~\ref{fig:relic}, we present the value of the coupling, $g$, that is required to generate a thermal relic abundance of $\Omega_{\phi_1} h^2 = 0.12$, as a function of $m_{\phi_1}$, and for $\Delta m = 0.1$~GeV and $f_1 = 0$. The results are shown for three different values of $f_2/g$. As the ratio $f_2/g$ increases, the $\phi_2 \phi_2$ annihilation cross section grows, necessitating a smaller value of $g$. In our regime of interest, $|g|, |f_2| \gg |f_{1}|$, a combination of coannihilations ($\phi_1 \phi_2 \rightarrow hh$, $\phi_1 \phi_2 \rightarrow f\bar{f}$), $\phi_2$ annihilations ($\phi_2 \phi_2 \rightarrow hh$, $\phi_2 \phi_2 \rightarrow f\bar{f}$), and $t$-channel $\phi_1$ annihilations ($\phi_1 \phi_1 \rightarrow hh$) govern the relic abundance.

We are now in a position to comment on the degree of fine tuning that is required in this model to be consistent with constraints from direct detection experiments. To obtain the required hierarchy, $|g|, |f_2| \gg |f_1|$, there must exist cancellations between the terms in Eq.~\ref{f1rel}. Comparing the values of $f_2$ and $g$ that are required to obtain the desired thermal relic abundance to the upper limits on $f_1$ shown in Fig.~\ref{fig:direct}, we find that these cancellations must occur at the roughly $\mathcal{O}(0.1-10\%)$ level. For example, in the case of $g=f_2$, such cancellations would require $f_1/g \lsim 10^{-3}$, 0.03, or 0.1, for $m_{\phi}=10 \, {\rm GeV}, 100 \, {\rm GeV}$, or 1 TeV, respectively. Alternatively, instead of a cancellation between terms, this hierarchy of couplings could arise if all three of $\cos 2 \theta$, ${\rm Im}(\eta)$, and $\kappa$ are small (or if all three of $\sin 2 \theta$, ${\rm Re}(\eta)$, and $\kappa$ are small). 
\section{Constraints From $\phi_2$ Decays}
\label{phi2decay}
In this model, $\phi_2$ can decay into a $\phi_1$ and Standard Model particles. These decays could impact the light element abundances or the ionization history of the universe, leading to nontrivial cosmological constraints on the parameters of this model.

The process $\phi_2 \to \phi_1 \gamma \gamma$ can proceed through an off-shell Higgs boson, with contributions arising from the interaction term proportional to $g$ in Eq.~\ref{eq:Lh-mass}, in combination with the effective coupling of the Higgs boson to photons, represented by $(c_{\gamma\gamma} h/v_h) F_{\mu\nu} F^{\mu\nu}$, where $F_{\mu\nu}$ is the electromagnetic field strength tensor. 
The width for this process is given by
\begin{align}
&\Gamma_{\phi_2 \rightarrow \phi_1 \gamma \gamma} = \frac{g^2 c^2_{\gamma \gamma} (\Delta m)^7}{210 \pi^3 m_h^4 m_{\phi_2}^2} \\
&\approx 4 \times 10^{-22} \, {\rm s}^{-1} \, \bigg(\frac{g}{0.3}\bigg)^2 \bigg(\frac{\Delta m}{ {\rm MeV}}\bigg)^7\bigg(\frac{\rm TeV}{m_{\phi_2}}\bigg)^2, \nonumber
\end{align}
where $c_{\gamma \gamma} \approx -2.03 \times 10^{-3}$  \cite{Ghosh:2019jzu}.   

For $\Delta m > 2m_e$, decays to $\gamma\gamma$ are always subdominant to decays to fermions.
In the  $2 m_f \ll \Delta m \ll m_{\phi_2}$ limit, the width for the process $\phi_2 \to \phi_1 f \bar f$ is given by
\begin{align}
&\Gamma_{\phi_2 \rightarrow \phi_1 f\bar{f}} = \frac{ n_c g^2  m_f^2 \,(\Delta m)^5}{240 \pi^3 m_h^4 m_{\phi_2}^2},
\end{align}
where $n_c=1 \,(3)$ for decays to leptons (quarks). For $2 m_e \ll \Delta m \ll 2 m_{\mu}$, this yields
\be
\Gamma_{\phi_2 \rightarrow \phi_1 ee}\approx  10^{-12} \, {\rm s}^{-1}  \bigg(\frac{g}{0.2}\bigg)^{\!2}
\bigg(\frac{\Delta m}{10 \, {\rm MeV}}\bigg)^{\!5} \bigg(\frac{\rm TeV}{m_{\phi_2}}\bigg)^{\!2} \!\! . ~~~~~
\ee
For larger mass splittings, $\Delta m \gg 2 m_{\mu}$, the decays proceed much more rapidly,
\be
\Gamma_{\phi_2 \rightarrow \phi_1 \mu\mu}\approx  400 \, {\rm s}^{-1} \, \bigg(\frac{g}{0.2}\bigg)^{\!2}
\bigg(\frac{\Delta m}{1 \, {\rm GeV}}\bigg)^{\!5 }\bigg(\frac{\rm TeV}{m_{\phi_2}}\bigg)^{\!2}. ~~~~
\ee
The total decay width further increases if we consider larger values of $\Delta m$, allowing other decay channels to become kinematically accessible.

Some of the most stringent constraints on long-lived particles follow from the ionization history of our universe and its impact on the cosmic microwave background. In particular, such constraints can exclude decays to electromagnetic final states in the following range~\cite{Slatyer:2012yq,Slatyer:2016qyl}:
\be
10^{13} {\rm s} \lsim \tau_{\phi_2} \lsim 10^{25}  \, {\rm s} \brac{\Delta m}{ m_{\phi_2}}.
\ee
In the regime of $\Delta m < 2 m_e$ and $m_{\phi} \gtrsim 100\,$GeV, the $\phi_2$ lifetime is long enough to avoid these constraints over nearly all of our parameter space. These constraints do, however, rule out a portion of the parameter space in which $2m_e \leq \Delta m \leq 2m_\mu$.

If $\phi_2$ decays produce photons or electrons with enough energy to disassociate helium nuclei (corresponding to $\Delta m \gsim 56.6 \, {\rm MeV}$), these reactions could potentially impact the primordial helium and deuterium abundances~\cite{Forestell:2018txr,Depta:2020zbh}. These considerations constrain the parameter space in this model with $\Delta m \gsim 60 \, {\rm MeV}$, $\tau_{\phi_2} \gsim 10^3 \, {\rm s}$, and $\Delta m/m_{\phi_2} \gsim 10^{-5}$, corresponding to the requirement, 
\be
\Delta m \gsim 5 \, {\rm GeV} \brac{0.3}{g}^{2/5} \biggl(\frac{ m_{\phi_2} }{{\rm TeV}}\biggr)^{2/5}.
\ee
These constraints become significantly weaker for $\tau_{\phi_2} \lsim 10^3 \, {\rm s}$. For $\Delta m \gsim {\rm GeV}$, these decays are expected to be safe from all cosmological bounds.

These constraints could be further relaxed if processes such as $\phi_2 f \rightarrow \phi_1 f$ are able to deplete the $\phi_2$ abundance before these particles can decay. If the rate for $\phi_2 \to \phi_1$ conversion is large compared to the Hubble rate, chemical equilibrium will be maintained between the $\phi_1$ and $\phi_2$ populations \cite{Baryakhtar:2020rwy,CarrilloGonzalez:2021lxm,Berlin:2023qco}. For $T \lsim \Delta m$, this would lead to the exponential suppression of the heavier population, $n_{\phi_2}/n_{\phi_1} \approx e^{-\Delta m/T}$. 
The cross section for the $\phi_2 \, f \rightarrow \phi_1 \,f$ process is given by~\cite{DiazSaez:2024nrq}
\be
\sigma_{\phi_2 f \to \phi_1 f} &=& \frac{g^2 m_f^2}{ 32 \pi E_{\phi_2} E_f |\vec  p_{i}| v \sqrt{s}} \biggl[ \log \brac{m_h^2 - t^-}{m_h^2 - t^+} 
\nonumber \\ 
&&~~~~
- \frac{4  |\vec  p_{i}||\vec  p_{f}| (m^2_h-4m_f^2
) }{(m^2_h-t^-)(m^2_h-t^+)} \biggr],
\ee
where $v$ is the relative velocity in the center-of-momentum frame, $\vec p_{i,f}$ are the initial and final momenta in this frame, $E_{\phi_2,f}$ are the particle energies, and 
$t^\pm$ is the mandelstam variable evaluated at $\cos \theta = \pm 1$, where $\theta$ is the scattering angle (for complete expressions, see Appendix~B of Ref.~\cite{DiazSaez:2024nrq}). To obtain the thermal average, we use Eq.~\ref{eq:gondolo} and evaluate the rate $\Gamma_{\phi_2 f \to \phi_1 f} = n_f \langle \sigma v\rangle_{\phi_2 f \to\phi_1 f}$, where $n_f \propto T^3$ is the number density of target fermions. Numerically, we find that for the range of mass splittings and couplings considered here, this rate can easily exceed the Hubble rate for $\Delta m \gsim T \gsim m_{\mu}$. 
Due to the small Yukawa coupling involved, the corresponding scattering rate with electrons is typically too small to maintain kinetic equilibrium. 
\section{Indirect Detection}
\label{ID}
DM particles annihilating in the Galactic Halo (or in other regions, such as within nearby dwarf galaxies) can lead to potentially observable fluxes of gamma rays~\cite{Fermi-LAT:2017opo,McDaniel:2023bju,DiMauro:2022hue} and cosmic-ray antimatter~\cite{Cuoco:2016eej,Giesen:2015ufa,Cholis:2019ejx}, allowing us to place constraints on the DM's annihilation cross section. Within the viable parameter space of this model, the relic abundance can be set by different combinations of coannihilations ($\phi_1 \phi_2 \rightarrow hh, ZZ, WW, f\bar{f}$), $\phi_2$ annihilations ($\phi_2 \phi_2 \rightarrow hh, f\bar{f}$), and $t$-channel $\phi_1$ annihilations ($\phi_1 \phi_1 \rightarrow hh$). Since only the last of these processes contributes to the $\phi_1 \phi_1$ annihilation rate, indirect detection signals will be highly suppressed if $m_{\phi} < m_h$. At higher masses, $m_{\phi} > m_h$, the magnitude of the annihilation rate depends on the mass and couplings of the DM, as well as on the mass splitting, $\Delta m$. 
For $m_\phi > m_h$, the cross section relevant for indirect detection is given by 
\be
    \langle \sigma v \rangle_{\phi_1 \phi_1 \rightarrow hh} &\approx& \frac{g^4 v_h^4}{64 \pi m^3_{\phi} (m^2_{\phi}-m^2_h)^{3/2}} \\
    &\sim&  10^{-26} {\rm cm}^3{\rm s}^{-1}  \,\, \bigg(\frac{g}{0.1} \bigg)^4 \bigg(\frac{150 \, {\rm GeV}}{m_\phi}\bigg)^6 \! , \nonumber
~~~~
\ee
where we have taken $f_1 \approx 0$.

A bright and statistically significant excess of GeV-scale emission has been detected from the region surrounding the Galactic Center~\cite{Goodenough:2009gk,Hooper:2010mq,Hooper:2011ti,Abazajian:2012pn,Hooper:2013rwa,Gordon:2013vta,Daylan:2014rsa,Zhou:2014lva, Calore:2014xka,Fermi-LAT:2015sau,Fermi-LAT:2017opo,Cholis:2021rpp,DiMauro:2021raz}. The spectrum, angular distribution, and overall intensity of this signal is consistent with that expected from DM in the form of an annihilating thermal relic. It was shown in Ref.~\cite{Hooper:2019xss} that DM annihilating to a pair of Higgs bosons with $\langle \sigma v \rangle_{\phi_1 \phi_1 \rightarrow hh} \sim 10^{-26} \, {\rm cm}^3/{\rm s}$ would provide a good match to this signal for masses in the range of $m_{\phi_1} \sim 125-150 \, {\rm GeV}$~\cite{Casas:2017rww, Hooper:2019xss}. In the same region of parameter space, DM annihilation could also produce the small excess of $\sim 10-20 \, {\rm GeV}$ antiprotons that has been observed in the cosmic-ray spectrum~\cite{Hooper:2019xss,Cuoco:2016eej,Cui:2016ppb,Cholis:2019ejx,Cuoco:2019kuu}. For these reasons, this model is well suited to generate a signal consistent with the observed features the Galactic Center Gamma-Ray Excess, as well as the excess of cosmic-ray antiprotons.
%
%
% 
%
%===============================================
\section{Electroweak Phase Transition}

\label{subsec:ewpt-gw}
The interaction between the Standard Model Higgs doublet and the complex singlet scalar field in this model can modify the Higgs potential at finite temperatures in a way that could potentially lead to a strong first-order EWPT in the early universe. Such a phase transition could play a central role in generating the observed baryon asymmetry through electroweak baryogenesis~\cite{Trodden:1998ym, Anderson:1991zb, Huet:1995sh, Morrissey:2012db},\footnote{For electroweak baryogenesis to be viable, the Higgs sector must also exhibit explicit or spontaneous CP-violation. In this study, we have assumed that no
CP-violation arises within the Higgs sector. However, this model could be straightforwardly extended, to include a $Z_2$ symmetric, CP-violating dimension-6 operator of the form, $y_t \, \overline{Q} \Tilde{H} \, \big( 1 + i c \frac{\phi \phi^{\star}}{\Lambda^2}\big) \, t_R \, + \, h.c.$, where $y_t$ is the top Yukawa coupling and $\Lambda$ is the cutoff scale~\cite{Cline:2012hg, Vaskonen:2016yiu, Grzadkowski:2018nbc, Ellis:2022lft, Roy:2025zvo}. 
}
and could result in the production of a stochastic background of gravitational waves.

A first-order EWPT can occur when the finite-temperature effective potential of the Higgs field develops at least two distinct local minima, separated by a potential barrier at temperatures above the critical temperature, $T_c$. At $T = T_c$, the two minima become degenerate in energy while remaining separated by this barrier. At one of these minima, electroweak symmetry remains unbroken, while it is broken at the other. As the universe cools below $T_c$, a global minimum in the broken-symmetry phase develops, characterized by non-zero field values and a lower potential energy. The system, initially trapped in the symmetric (false vacuum) phase, can then transition into the broken (true vacuum) phase via quantum or thermal tunneling~\cite{Linde:1977mm,Linde:1978px,Linde:1981zj}.

The transition into the true vacuum proceeds through the nucleation of bubbles in the broken phase within the symmetric thermal plasma. Once formed, these bubbles grow and eventually coalesce, thereby converting the remaining volume of space into the broken phase.  The bubble nucleation rate per unit volume at finite temperature, in the semiclassical approximation, satisfies the following~\cite{Langer:1969bc,Coleman:1977py,Affleck:1980ac}:
\be
\Gamma_B \propto T^4 \exp\left( -\frac{ S_3(T) } {T} \right) ,
\ee
where $S_3(T)$ is the three-dimensional Euclidean action, as evaluated at the bounce configuration.
%, which satisfies the classical equations of motion. 
For the phase transition to successfully complete, the nucleation rate must be sufficiently large to produce at least one bubble per Hubble volume, per Hubble time, as realized when $S_3(T) \sim 140 \, T$~\cite{Linde:1981zj, Mazumdar:2018dfl, Quiros:1999jp}. The corresponding nucleation temperature, $T_n \,(\lesssim T_c)$, 
is the highest temperature at which this condition is 
satisfied as the universe cools.
In our analysis, we compute the bounce solution and corresponding action using the path deformation algorithm, as implemented in the publicly available package, \cosmotransitions~{\tt (v2.0.6)}~\cite{Wainwright:2011kj}.

To facilitate successful electroweak baryogenesis, the electroweak sector must undergo a strong first-order phase transition, corresponding to the following criterion:
\begin{equation}
    \xi_n= \frac{v_h(T_n)}{T_n} > 1 \, ,
\end{equation}
where $v_h(T_n)$ is the vev along the Standard Model Higgs field direction evaluated at the nucleation temperature, $T_n$. This condition ensures that any baryon asymmetry that is generated will not be washed out after the phase transition has ended. 

To analyze the thermal evolution of the effective potential, one begins with the tree-level scalar potential at zero temperature, $V_0(h, \phi_1, \phi_2)$. From Eqs.~\ref{Vsm}–\ref{grel}, this potential can be expressed as
% %
 \begin{align}
 \label{Vtree}
 V_0 = &\, -\frac{1}{2} \mu_h^2 h^2 + \frac{1}{4}\lambda_h h^4 + \frac{1}{2}\left(m_0^2 + \frac{1}{2} \kappa h^2 \right)(\phi_1^2 + \phi_2^2 ) \nonumber\\ & + \frac{1}{2}\left(\rho_{0{_R}}^2 + \frac{\eta_R}{2} h^2\right)(\phi_1^2 - \phi_2^2) - \left(\rho_{0{_I}}^2 + \frac{\eta_I}{2} h^2 \right)\phi_1 \phi_2  \nonumber\\ & + \frac{1}{4} \left( \lambda_1 \phi_1^4 +  \lambda_2 \phi_2^4 + \lambda_{12} \phi_1^2 \phi_2^2 \right) \,,
 \end{align}
% %
where we have used the shorthand notation,  
\be
\eta_R &=& \mathrm{Re}(\eta) ~,~~~~~   \eta_I  = \,\mathrm{Im}(\eta) \nonumber \\
\rho_{0{_R}}^2 &=& \mathrm{Re}(\rho_{0}^2) ~,~~~  \rho_{0{_I}}^2 =  \mathrm{Im}(\rho_{0}^2).
\ee
Here, we include quartic self-interactions for the singlet fields $\phi_1$ and $\phi_2$, parameterized by $\lambda_1$, $\lambda_2$, and $\lambda_{12}$. In the special case $\lambda_1 = \lambda_2 = \lambda_{12}$, the scalar sector exhibits an enhanced approximate $U(1)$ symmetry. Allowing these couplings to differ explicitly breaks this symmetry and provides additional freedom in shaping the scalar potential.
In particular, the mixed quartic coupling $\lambda_{12}$ controls the interaction between the two singlet directions and plays an important role in determining the structure of the potential barrier and the resulting phase transition pattern.
We note that, for generic values of $\lambda_1$, $\lambda_2$, and $\lambda_{12}$, the singlet sector exhibits a $Z_2 \times Z_2$ symmetry under independent sign flips of $\phi_1$ and $\phi_2$. The $Z_2$ symmetry relevant for dark matter stability corresponds to the diagonal subgroup under which both fields transform simultaneously. The approximate $U(1)$ symmetry is already broken by the mass and Higgs portal terms, and allowing the quartic couplings to differ introduces additional explicit breaking without affecting the consistency of the model.

In this study, we will work in the limit of $\sin 2\theta = 0$, which implies $\rho_{0_{I}}^2 = -v_h^2 \,\eta_I /2 $. 
At zero temperature, $V_0$ is minimized for 
\be 
\langle h \rangle = v_h \, , ~~ \langle \phi_1 \rangle = 0 \, ,~~ \langle \phi_2 \rangle = 0~.
\ee
The following quantities can be treated as the free parameters of the scalar potential given in Eq.~\ref{Vtree}:
\be
\label{free-para}
m_{\phi_1} \, , \, \Delta m \, , \, f_1 \>  , \> f_2 \, , \, g \, , \,\lambda_1 \, , \, \lambda_2 \, , \, \lambda_{12}  \> .
\ee
%where $\Delta m \equiv m_{\phi_2}-m_{\phi_1}$.
Using Eqs.~\ref{f1rel}-\ref{grel},  the other parameters of the potential can be expressed as
\begin{align}
%\label{relations}
\eta_I &= - g \,, \quad \kappa = f_2 \, , \quad \eta_R = -\kappa \, , \label{rho0Rsq} \\
m_0^2 &= \frac{1}{2}(m_{\phi_1}^2 + m_{\phi_2}^2 - \kappa v_h^2) \, , \nonumber \\
\rho_{0_R}^2 &= \frac{1}{2}(m_{\phi_1}^2 - m_{\phi_2}^2 
+ \kappa v_h^2 
) \, ,  \nonumber
\end{align}
where we have taken $f_1 \approx 0$. The zero-temperature, tree-level potential given in Eq.~\ref{Vtree} receives quantum corrections from all of the Standard Model degrees-of-freedom that couple to the $h$, $\phi_1$, or $\phi_2$ fields. A discussion of these corrections, as well as the finite-temperature corrections, is provided in Appendix~\ref{subsec:effpot}.

Since the Standard Model predicts a crossover EWPT, we must rely on interactions bewteen the the Higgs and $\phi_{1,2}$ fields to obtain a strong first-order phase transition. The coupling, $f_1$, is tightly constrained by direct detection experiments and must be quite small (see Fig.~\ref{fig:direct}). In contrast, $f_2$ and $g$ are not directly involved in DM-nucleon scattering at tree level and can be much larger. 
The impact of $g$ on the Higgs thermal corrections is generally mild compared to that of $f_2$, as seen in Eq.~\ref{Eq:DaisyCoeffs} of Appendix~\ref{subsec:effpot}.
For these reasons, it is $f_2$ that generally plays the most important role in facilitating a strong first-order phase transition.

The thermal evolution of the scalar potential in this model can proceed through a different series of events,  
depending on the parameter values that are selected. In addition to the standard, one-step first-order phase transition, it is possible that this phase transition may have taken place in two distinct steps. In such a scenario, the fields $\phi_1$ and $\phi_2$ first acquire non-zero vevs through a second- or first-order phase transition. This is followed by a first-order phase transition in which the Higgs field acquires a non-zero vev while $\phi_1$ and $\phi_2$ eventually return to the symmetric phase with vanishing vevs.

The residual $Z_2$ symmetry under which $\phi_i \to -\phi_i$ ensures the stability of the DM and forbids non-zero vevs for either field at zero temperature. 
At high temperatures, however, these fields can temporarily develop non-zero vevs due to thermal effects. The pattern of symmetry breaking in the singlet sector can be understood by examining the coefficients of their quadratic terms in the scalar potential, which can be derived from the field-dependent mass relations given in Eq.~\ref{mH22-masssq}.
At finite temperature, prior to electroweak symmetry breaking, and in the limit $\lambda_{12} \approx 0$ and $\rho_{0_I}^2 \approx 0$, the following relations hold:
\be 
g \approx 0 \, ,~~~~~
m^2_{\phi_{1,2}} \approx m_0^2 \pm \rho_{0_R}^2 \, .~~~~~~
\ee
So, for $\rho_{0_R}^2 > m_0^2 > 0$,  $\phi_2$ can acquire a non-zero vev, while $\phi_1$ remains at the origin.
 Thus, depending on the sign and magnitude of $\rho_{0_R}^2$, only one of the singlet fields acquires a non-zero vev.
When $\rho_{0_I}^2$ (or equivalently, $g$) is non-zero, the $\rho_{0{_I}}^2 \phi_1 \phi_2$ term of the potential can result in a tadpole contribution which can lead to a non-zero vev for the other field at finite temperature.
At sufficiently high temperatures, however, large positive thermal mass corrections (see Eq.~\ref{Eq:DaisyCoeffs} of Appendix \ref{subsec:effpot}) can cause the coefficients of the quartic field terms to become positive, restoring the $Z_2$ symmetry.

\begin{table}[t!]
\setlength{\tabcolsep}{2pt}
   \centering
\small{\begin{tabular}{|c|c|c|c|c|c|c|}
\hline
\Tstrut
Input/Observables & \makecell{BP1} & \makecell{BP2}\\
\hline
\Tstrut
$\lambda_1$ & 0.4  & 1.6  \\[0.08cm]
$\lambda_2$ & 0.4  & 1.3 \\[0.08cm]
$\lambda_{12}$ & 0.3 & 0.2  \\[0.08cm]
$f_1$ & $0.0004$ & $0.0009$   \\[0.08cm]
$f_2$ & 0.26 & 0.49 \\[0.08cm]
$g$  & 0.046 & 0.16\\[0.08cm]
\hline
\Tstrut
$m_{\phi_1}$~(GeV) & 68.7 & 130  \\[0.08cm]
$m_{\phi_2}$~(GeV) & 71.4 & 150  \\[0.1cm]
\hline
\Tstrut
$\Omega_{\phi_1} h^2$ & 0.12 & 0.12  \\[0.08cm]
$\sigma^{\rm SI}_{\phi_1 N}$~(cm$^2$) & $1.4\times 10^{-48}$ & $1.8\times 10^{-48}$     \\[0.15cm]
$\langle \sigma v\rangle_{\phi_1 \phi_1} $~(cm$^3/$s)  & $ 4.9\times 10^{-30}$ & $ 1.6\times 10^{-26}$   \\ [0.1cm]
\hline
\end{tabular}
}
\caption{The parameters for two selected benchmark models (BP1 and BP2), along with the masses of the dark matter and its excited state, the dark matter's thermal relic abundance, and the cross sections relevant for direct and indirect detection.}
\label{tab:BPs}
\end{table}

The parameter, $\rho_{0_R}^2$, depends on the mass splitting, $\Delta m$, for a fixed value of $f_2$, making $\Delta m$ a key driver of symmetry breaking in the singlet sector.
For some values of $\Delta m$, the singlet fields never acquire non-zero vevs at high temperatures, resulting in a one-step phase transition along the Higgs field direction. Conversely, in order for a two-step phase transition to occur, $\Delta m$ and $g$ must be appropriately tuned to allow $\phi_1$ and $\phi_2$ to develop non-zero vevs. The nature of this phase transition along the singlet direction, whether second- or first-order, depends on the strength of thermally induced cubic terms.

\begin{table}[t!]
\renewcommand{\arraystretch}{1.8}
\setlength{\tabcolsep}{1.7pt}
\centering
\scriptsize{
\begin{tabular}{|c|c|c|c|c|c|c|c|c|}
\hline
 & \multicolumn{2}{c|}{$T_i$ (GeV)} & $\{h, \phi_1, \phi_2\}$ & \multirow{2}{*}{$\xrightarrow[\text{type}]{\text{PT}}$} & $\{h, \phi_1, \phi_2\}$  & \multirow{2}{*}{$\xi_n$} & \multirow{2}{*}{$\alpha$} & \multirow{2}{*}{$\beta/H_n$} \\
\cline{2-3}
 & $T_c$ & $T_n$ & (False) (GeV) & & (True) (GeV) & & & \\
\hline
\multirow{2}{*}{BP1} 
& 343 & 343 & \{0, 0, 0\} & S & \{0, 1, 9\} &  &  &  \\   
\cline{2-6}
& 90 & 69 & \{0, 29, 170\} & F & \{238, 0, 0\} & 3.4 & 0.13 & 420 \\
\hline
\multirow{2}{*}{BP2} 
& 257 & 257 & \{0, 0, 0\} & S & \{0, 1, 5\} &  &  &  \\   
\cline{2-6}
& 121 & 109 & \{0, 21, 105\} & F & \{203, 0, 0\} & 1.9 & 0.035 & 2507 \\
\hline
\end{tabular}
}
\caption{
The characteristics of the phase transitions predicted in our benchmark models, BP1 and BP2, as defined in Table~\ref{tab:BPs}. Both of these models predict a  second-order phase transition (S), which is followed by a first-order transition (F), the latter of which triggers electroweak symmetry breaking. Listed in this table are the values of the critical temperature, $T_c$, the nucleation temperature, $T_n$, and the field values at the false and true vacua. The values of $\xi_n$, $\alpha$, and $\beta/H_n$ are also provided in each case.
}
\label{TcTn}
\end{table}
\section{Benchmark Scenarios}
\label{sec:benchmarks}
In Table~\ref{tab:BPs}, we present parameter values for two representative benchmark models, BP1 and BP2. For BP1, $m_{\phi_1}$ lies below $m_h$ but above $m_h/2$, while for BP2, $m_{\phi_1} > m_h$.
These specific choices for 
$m_{\phi_1}$ have distinct implications for DM detection.

For both benchmarks, the predicted spin-independent direct detection cross sections 
lie well below current bounds from LUX-ZEPLIN, while remaining within the projected 
sensitivity of next-generation experiments.

The indirect detection prospects, however, differ markedly between the two cases. 
For BP1, the condition $m_{\phi_1} < m_h$ kinematically forbids annihilation into 
Higgs boson pairs, resulting in 
$\langle\sigma v\rangle_{\phi_1\phi_1} \simeq 4.9 \times 10^{-30}~\mathrm{cm}^3/\mathrm{s}$, 
which is far below the sensitivity of current gamma-ray and cosmic-ray searches. 
In contrast, for BP2 with $m_{\phi_1} = 130~\mathrm{GeV}$, the 
$\phi_1\phi_1 \rightarrow hh$ channel is open and dominates the annihilation rate, 
yielding $\langle\sigma v\rangle_{\phi_1\phi_1} \simeq 1.6 \times 
10^{-26}~\mathrm{cm}^3/\mathrm{s}$. This value lies in the range required to explain 
the Galactic Center Gamma-Ray Excess.
A closely related scenario with $m_{\phi_1} \simeq 130~\mathrm{GeV}$ and dominant 
$\phi_1 \phi_1 \rightarrow hh$ annihilation has been studied in detail in 
Ref.~\cite{Casas:2017rww}, where it was shown to provide an excellent fit to the 
Galactic Center Excess spectrum, with a reported $p$-value of $\sim 0.6$--$0.7$. 
Independently, it has been shown that annihilation cross sections of this size 
into hadronic final states are compatible with the AMS-02 antiproton data 
\cite{Cuoco:2016eej, Cholis:2019ejx, Hooper:2019xss}. 
Given the close correspondence between that setup and our BP2, we do not repeat 
the full likelihood analysis here, but instead rely on these established results 
for this mass range and annihilation channel.

Both benchmarks exhibit a two-step phase transition, whose numerical characteristics are given in Table~\ref{TcTn}.
In the first step, the singlet fields acquire non-zero vevs, spontaneously breaking the $Z_2$-symmetry.
Subsequently, at a lower temperature, a second phase transition breaks electroweak symmetry. This latter transition satisfies $\xi_n >1$, which is a key requirement of electroweak baryogenesis. At the end of the second step, the singlet fields transition back to a state with zero vevs, thereby restoring the $Z_2$-symmetry which ensures the stability of the DM.

%%%%%%%%%%%%%%%%%%
\begin{figure*}[t]
    \centering
    \includegraphics[width=0.49\linewidth]{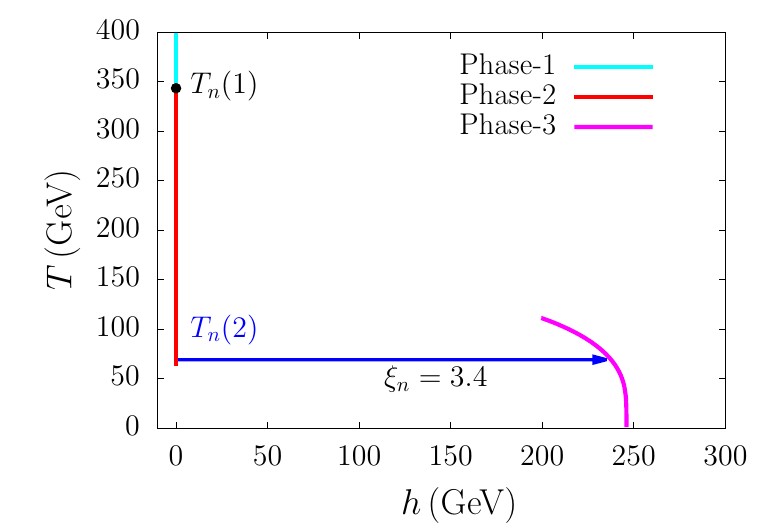} \,\,\,
    \includegraphics[width=0.49\linewidth]{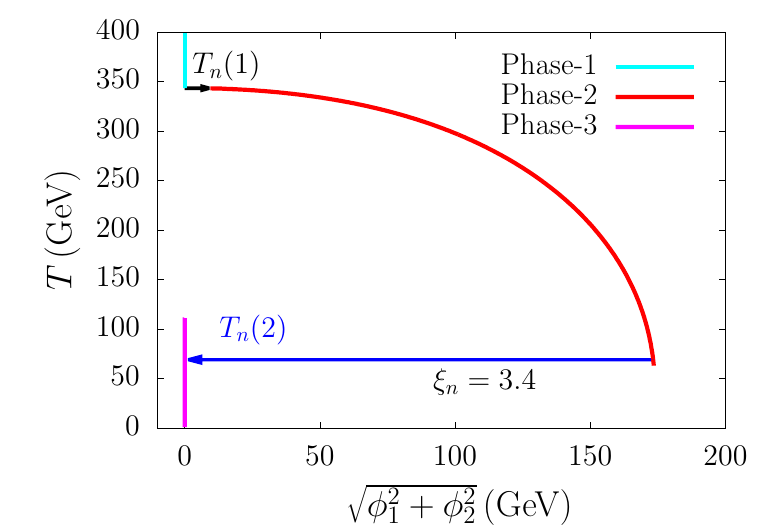}
    \caption{The evolution of the field values in the case of our benchmark model, BP1. Each color represents a distinct minimum of the potential, and the lines describe the phase evolution along different field directions as a function of temperature.
The left and right plots illustrate BP1's phase evolution along the Standard Model Higgs field, $h$, and the singlet direction, $\sqrt{\phi_1^2 + \phi_2^2}$, respectively. 
Arrows indicate the transition path from the false vacuum to the true vacuum, as calculated at the nucleation temperature, $T_n$. The black point in the left plot highlights the transition during which the corresponding field values exhibit minimal variation. 
}
\label{phasediag}
\end{figure*}

In Fig.~\ref{phasediag}, we plot the 
field values at various minima for the case of BP1. Each color in this figure represents a distinct minimum of the potential, and the lines depict phase evolution along different field directions as a function of temperature. The two-step transition involves three distinct phases, as indicated in Table~\ref{TcTn}:
\begin{itemize}
    \item Phase 1 (cyan line): The symmetric phase, in which all of the vevs are equal to zero.
    \item Phase 2 (red line):  The $Z_2$ symmetry is broken ($\phi_1, \phi_2 \neq 0$), but electroweak symmetry remains unbroken ($h = 0$).
    \item Phase 3 (magenta line): The phase in which electroweak symmetry is broken ($h \neq 0$), but the $Z_2$ symmetry is restored ($\phi_1 = \phi_2 = 0$).
\end{itemize}

At high temperatures, the system starts in the symmetric phase with all scalar fields at the origin. As the universe cools, a second-order transition occurs at $T=343$~GeV and the singlet fields develop non-zero vevs as the system moves to Phase 2. In the left frame, the black point between Phase 1 and Phase 2 indicates that the field, $h$, remains unchanged during this transition, while the black arrow in the right plot shows that $\phi_1$ and $\phi_2$ develop non-zero vevs.
At a lower temperature, the electroweak symmetry-breaking minimum (Phase 3) develops.
The blue arrow marks the transition path at the nucleation temperature, $T_n = 69$~GeV, where the system tunnels from Phase 2 to Phase 3. The value of $h$ reaches 246 GeV at $T=0$.

This sequence of events 
has important cosmological implications.
In the parameter space featuring a two-step phase transition, the $Z_2$ symmetry is spontaneously broken by thermal effects when $\phi_1$ and $\phi_2$ acquire non-zero vevs, leading to the formation of potentially problematic domain walls.
If these domain walls were to persist to later times, they would come to dominate the energy density of the universe and spoil the success of the standard cosmological model.
During the second step of our phase transition, however,  
$\phi_1$ and $\phi_2$ return to the origin of field space, resulting in the destruction of these domain walls at early times~\cite{Cline:2012hg,Espinosa:2011eu}.

In some region of parameter space, the nucleation of electroweak bubbles within domain walls could accelerate the phase transition process~\cite{Blasi:2022woz, Agrawal:2023cgp}, potentially modifying the nucleation temperature, $T_n$, and the resulting gravitational wave spectrum. A detailed study of these effects is beyond the scope of this work and is left for future investigation. Here, we assume that the bubbles grow homogeneously and neglect the impact of any domain walls on the resulting phase transition.
\section{Gravitational Waves}
\label{sec:gravitationalwaves}
A first-order phase transition could result in the generation of a potentially observable stochastic background of gravitational waves. Such a signal would represent a way to probe the dynamics of the phase transitions that took place in the early universe.      
The theoretical framework for estimating the amplitude of gravitational waves from a first-order phase transition is described in Appendix~\ref{GWs_section}.

The key parameters that govern the properties of this signal are
\be
\label{eq:keypara}
\alpha \,,~~ T_n \, ,~~\beta/H_n \, , ~~ v_w \, , 
\ee
where $\alpha = \Delta \rho/\rho_{\rm rad}$ is the ratio of the latent heat released to the radiation density, $T_n$ is the nucleation temperature, $\beta/H_n$ is the ratio of the bubble nucleation rate to the Hubble rate evaluated at $T_n$, and $v_w$ is the wall velocity. The parameter values used in our benchmark models 
are summarized in Table~\ref{TcTn}.
Here, we consider $v_w \sim 1$, assuming that the expanding bubbles attain a relativistic terminal velocity in the plasma.

 As detailed in Appendix~\ref{GWs_section}, the spectrum of the resulting gravitational waves receives contributions from three primary sources: 1) collisions of expanding bubbles, 2) sound waves from bubbles expanding through the plasma, and 3) magneto-hydrodynamic (MHD) turbulence. The peak frequency and overall amplitude of the resulting gravitational wave spectrum are primarily determined by the energy density and wavelength of the sound waves.

\begin{figure}[t!]
\begin{center}
    \centering
\includegraphics[width=1.02\linewidth]{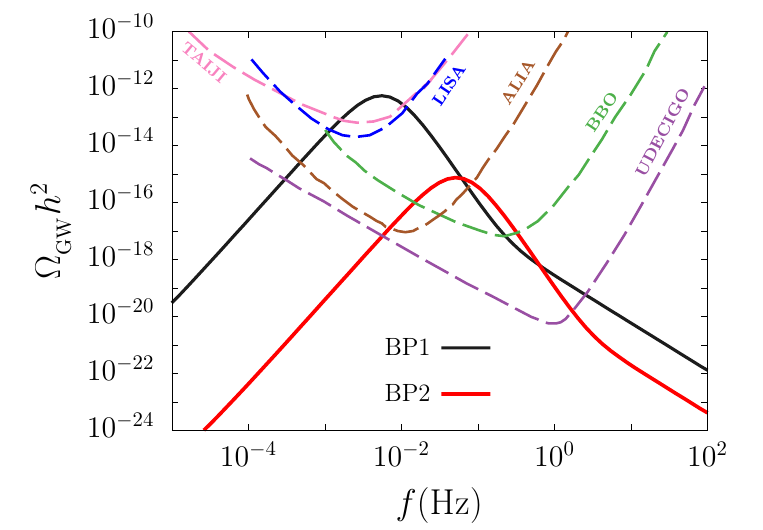}
\caption{The gravitational wave energy density spectra predicted for our benchmark models, BP1 and BP2. These results are compared to the projected sensitivities of the proposed gravitational wave detectors LISA, Taiji, ALIA, BBO, and U-DECIGO. 
}
\label{fig:GWs-plot}
\end{center}
\vspace{-0.5cm}
\end{figure}

The gravitational wave spectra that result in our benchmark models, BP1 and BP2, as calculated using Eqs.~\ref{soundgw}--\ref{eq:h-star}, are shown in Fig.~\ref{fig:GWs-plot}. These predictions are compared to 
the projected sensitivities of several proposed space-based detectors, including LISA~\cite{LISA:2017pwj}, ALIA~\cite{Gong:2014mca}, Taiji~\cite{Hu:2017mde}, BBO~\cite{Corbin:2005ny}, and U-DECIGO~\cite{Kudoh:2005as}.

To quantify detectability, we evaluate the signal-to-noise ratio (SNR), defined as~\cite{Caprini:2015zlo, Babak:2021mhe, Smith:2019wny}
\begin{equation} \small
\text{SNR} = \sqrt{\delta \, \mathcal{T} \int_{f_{\text{min}}}^{f_{\text{max}}} df \left[\frac{h^2 \Omega_{\text{GW}}(f)}{h^2 \Omega_{\text{exp}}(f)}\right]^2 } \,,
\label{snr}
\end{equation}
where $\delta$ denotes the number of independent channels (e.g.\ $\delta = 1$ for LISA and $\delta = 2$ for BBO and U-DECIGO), and $\mathcal{T}$ is the mission duration, taken here to be $\mathcal{T} = 5$ years. The quantity $h^2 \Omega_{\text{exp}}(f)$ represents the effective noise power spectral density of the experiment.
A GW signal is considered detectable if the SNR exceeds a threshold value that depends on the detector configuration. For LISA, this threshold is typically $\text{SNR} \sim 50$ for a four-link design and $\sim 10$ for a six-link configuration~\cite{Caprini:2015zlo}.

While we compute the SNR explicitly for LISA, which serves as the standard benchmark for space-based GW detectors with well-defined sensitivity and detection criteria, the situation is less clear for future experiments such as BBO and U-DECIGO. For these detectors, the precise SNR thresholds depend on the specific mission configuration and data analysis strategy, and are not yet universally established. 
For this reason, we primarily use LISA to quantify detectability through the SNR, while for other proposed experiments we rely on a qualitative comparison between the predicted GW spectra and their projected sensitivity curves.

With this in mind, we find $\text{SNR} = 29.2$ for BP1 and $\text{SNR} = 0.005$ for BP2 in the case of LISA. Taking a representative detection threshold of $\text{SNR} = 10$, BP1 is potentially observable, whereas BP2 is not detectable by LISA, as can be seen from Fig.~\ref{fig:GWs-plot}, where the GW spectrum of BP2 lies well below the LISA sensitivity curve over the entire frequency range, and therefore does not enter the detectable region.
Nevertheless, BP2 may still lie within the sensitivity reach of more advanced future detectors such as BBO and U-DECIGO, as indicated by the comparison with their projected sensitivity curves.

\section{Collider Production}
\label{collider}
Probing the physics of the Higgs sector 
is a key priority for existing and future high-energy colliders, including the LHC \cite{Cepeda:2019klc}, FCC-ee \cite{Agapov:2022bhm}, FCC-hh \cite{Benedikt:2022kan}, ILC \cite{ILCInternationalDevelopmentTeam:2022izu}, and a muon collider~\cite{Accettura:2023ked}. An important part of this program are measurements of the invisible width of the Higgs, as well as other exotic decay channels~\cite{Curtin:2013fra}.

%%%%%%%%%%%%%%%%%%
%
%%%%%%%%%%%%
%

For $m_{\phi_1}, m_{\phi_2} \lesssim m_h/2$, the dominant new Higgs decay channels in this model are $h \to \phi_1 \phi_2$ and $h \to \phi_2 \phi_2$, which arise from the couplings $g$ and $f_2$, respectively ($h \to \phi_1 \phi_1$ is suppressed by the small value of $f_1$ required to satisfy direct detection constraints). Note that the value of $g$ also governs the DM's thermal relic abundance in the coannihilation limit, allowing us to place a lower limit on the branching fraction to $\phi_1 \phi_2$ from relic abundance considerations. The upper bound on the Higgs invisible branching fraction from the LHC ($\lesssim 0.11$)~\cite{ATLAS:2022yvh,CMS:2023sdw,ATLAS:2023tkt} constrains a substantial portion of the $m_{\phi} < m_h/2$ parameter space in this model~\cite{Ghorbani:2014gka, Casas:2017jjg, DiazSaez:2024nrq}.

For heavier DM, $\phi_1$ and $\phi_2$ can be produced through off-shell, Higgs mediated processes,
\begin{align}
    p p \to h^{\star} X \to  \phi_1  \phi_2 X, \\ 
    p p \to h^{\star} X \to  \phi_2  \phi_2 X,\nonumber
\end{align}
where $X$ represents an associated Standard Model object, such as a hard initial-state jet or photon. Again, the production of $\phi_1 \phi_1$ is suppressed by the small value of~$f_1$.

Although decays of the form $\phi_2 \rightarrow \phi_1 + {\rm SM}$ could lead to distinctive signatures, the lifetime for this process is too long in much of the parameter space under consideration to be observed at colliders. If the mass splitting is sufficiently large, however, the $\phi_2$ lifetime could be short enough to be observed at the LHC~\cite{Lee:2018pag}, or at future experiments such as MATHUSLA~\cite{Curtin:2018mvb}, as discussed in Ref.~\cite{DiazSaez:2024nrq}. Even larger mass splittings could be probed at the LHC by focusing on final states with a hard jet/photon, large transverse missing energy, and additional Standard Model particles from $\phi_2$ decays~\cite{Goncalves:2025snm}.
\section{Conclusions}
\label{Conclusions}
Minimal scalar dark matter with a Higgs portal coupling offers a compelling and highly predictive scenario for thermal freeze out. This model, however, has been almost entirely excluded by direct detection experiments. 
In this paper, we have presented a simple and viable variation of this scenario in which a {\it complex} scalar field couples to the Higgs portal through operators that explicitly break this field's global $U(1)$ symmetry. In the mass basis, the complex field decomposes into two non-degenerate eigenstates, $\phi_1$ and $\phi_2$. The heavier of these two states is unstable, while the lighter state can serve as a dark matter candidate. In selected regions of parameter space, the leading Higgs portal interaction can be off-diagonal, leading to inelastic scattering with nuclei, and greatly relieving the tension with direct detection experiments.

The thermal relic abundance of dark matter in this model is governed by both annihilation and coannihilation processes. We have identified regions of parameter space in this model with $\phi_1 \sim 125-150 \, {\rm GeV}$ that predict a signal which is consistent with the observed features of the Galactic Center Gamma-Ray Excess.

The interactions of $\phi_1$ and $\phi_2$ in this model modify the shape of the Higgs potential, potentially facilitating a strong first-order electroweak phase transition. 
In addition to being necessary for successful electroweak baryogenesis, such a transition could lead to a potentially observable background of gravitational waves. 
In the minimal Higgs portal model, a strong first-order electroweak phase transition would require a value of the portal coupling that is ruled out by direct detection constraints. In contrast, large portal couplings are phenomenologically viable in this model, making it possible to modify the Higgs potential in a way that could allow for a strong first-order phase transition.

\bigskip
{\bf Acknowledgments.} 
We thank Paddy Fox and Carlos Wagner for helpful conversations.
DH is supported by the Office of the Vice Chancellor for Research at the University of Wisconsin-Madison, with funding from the Wisconsin
Alumni Research Foundation. Fermilab is operated by the Fermi Research Alliance, LLC under Contract No.~DE-AC02-07CH11359 with the U.S. Department of Energy, Office of Science, Office of High Energy Physics. DR is supported by the U.S. Department of Energy, Office of Science, Office of Workforce Development for Teachers and Scientists, Office of Science Graduate Student Research (SCGSR) program. The SCGSR program is administered by the Oak Ridge Institute for Science and Education for the DOE under contract number DE‐SC0014664. SR is supported by the U.S.~Department of Energy under contracts No.\ DEAC02-06CH11357 at the Argonne National Laboratory.  SR would like to thank the University of Chicago, Fermilab, the Aspen Center for Physics, and the Perimeter Institute for Theoretical Physics, where a significant part of this work was carried out.
\bibliographystyle{utphys3}
\bibliography{biblio}
%
%%%%%%%%%%%%%%%%%%%%%%%%%%%%%%%%%%%%%%%%%%%%%%%%%%%%%%%%%%%%
%\appendix
\appendix
\label{appendix}
%
%%%%%%%%%%%%%%%%%%%
\section{Direct Detection at One Loop}\label{oneloopDD}
%%%%%%%%%%%%%%%%%%
The tree-level direct detection cross section for elastic, spin-independent, $\phi_1 N \to \phi_1 N$ scattering is governed by $f_1$, which is chosen to be small in our benchmark scenarios. There are loop corrections to this process, however, which depend instead on $g$ and/or $f_2$. 

The main loop diagrams that contribute to this process include a triangle correction to the $\phi_1 \phi_1 h$ vertex. For the process $\phi_1(p) N(k) \to \phi_1(p^\prime) N(k^\prime)$, the one-loop vertex contribution, $\delta f_1$, is given by
\be
2 \delta f_1 v = (-igv)^2 (I_1 + I_2),
\ee
where the integrals corresponding to the two scalar triangles are 
\be
I_1 &\approx& -\frac{3 i m_h^2}{v} \int \frac{d^4\ell}{(2\pi)^4} \frac{1}{( \ell^2 - m_h^2)^2 [ (\ell + p)^2 - m_{\phi_2}^2]},~~~~~ \\
I_2 &\approx& -2if_2 v \int \frac{d^4\ell}{(2\pi)^4} \frac{1}{( \ell^2 - m_{\phi_2}^2)^2 [ (\ell + p)^2 - m_h^2]},~~~~~
\ee
where $q = p-p^\prime$ is the non-relativistic momentum transfer, and we have dropped terms of order $q^2/m_h^2$ and $q^2/m_{\phi_2}^2$. 
Using dimensional regularization, the vertex correction is 
\be
2 \delta f_1 v = -\frac{ i(gv)^2 }{32\pi^2} \left( 
\frac{3  m_h^2}{v} \Delta_1 + 2f_2 v \Delta_2
\right),
\ee
where we have defined the Feynman parameter integrals, 
\be
\Delta_1 &=& \int_0^1 dx  \frac{  2(1-x) }{   
 x^2 m_{\phi_1}^2
+ (1-x) m_{\phi_2}^2  - x ( m_{\phi_1}^2 - m_h^2)
} ,~~~~~ \\
\Delta_2 &=& \int_0^1 dx \,  \frac{  2( 1-x) }{
 x^2m_{\phi_1}^2
+ (1-x) m_{h}^2  - x ( m_{\phi_1}^2 - m_{\phi_2}^2)
}, \, 
\ee
and the full amplitude for this process is 
\be
i{\cal M} \approx \frac{ 2v(f_1 + \delta f_1) y_N  }{ m_h^2} [\bar u(k^\prime)u(k)] \, ,
\ee
where $y_N \sim m_N/v$ is the effective Higgs-nucleon coupling and $u$ is a four-component spinor.
Since we require $f_1 \ll 1$, the loop contribution to this process can be written as
\be
\sigma_{\phi_1 N}^{\rm SI, loop} \approx \frac{ |\delta f_1|^2 y_N^2 m_N^2   }{ \pi m_h^4 m_{\phi_2}^2 }\, .
\ee
We have numerically verified that the parameter choices for our benchmarks in Table \ref{tab:BPs}
are safe from direct detection bounds, predicting values of order $\sigma_{\phi_1 N}^{\rm SI, loop} \sim 10^{-49}$ cm$^2$.

For completeness, we note that there is also a box diagram that enables $\phi_1 N \to \phi_1 N$ scattering through double-Higgs exchange. We have numerically verified that this contribution is subdominant to the (already small) contributions from the vertex loop, and we do not consider it further in our analysis.  
\section{Higgs Potential at Finite-Temperature}

\label{subsec:effpot}
%%%%%%%%%%%%%%%%%%%%%

The one-loop radiative corrections to the zero-temperature tree-level potential, defined in Eq.~\ref{Vtree}, can be expressed in terms of the well-known Coleman-Weinberg (CW)~\cite{Coleman:1973jx} potential. In the $T = 0$ limit, this takes the following form:
\be
\label{VCW}
V_{\rm CW}(m_i^2) = \sum_{i} (-1)^{2s_i}\frac{  n_i m_i^4}{64\pi^2}  \left(\ln \frac{m_i^2}{Q^2}-C_i\right) ,
\ee
where $m_i \equiv m_i (h, \phi_1, \phi_2)$, $Q$ is the renormalization scale (which we take to be the electroweak scale, $Q = v_h$), and the sum is performed over all the particles in the model. The constants, $C_i$, depend on the choice of the renormalization scheme. In this work, we adopt the $\overline{\text{MS}}$ on-shell scheme, for which 
$C_{W^\pm,Z}=5/6$ and $C_{i}=3/2$ for all other particle species.
Here, $m_i$ is the tree-level field-dependent mass of the $i$-th species, and $n_i$ and $s_i$ are their associated numbers of degrees-of-freedom and spin, respectively.

The one-loop CW correction to the tree-level potential alters the masses and mixing angles of the scalar degrees-of-freedom and shifts the position of the electroweak minimum in field space.
To keep them fixed at tree-level, we modify the $\overline{MS}$ CW potential by adding suitable counterterms, $V_{\text{CT}}$, to the potential, which satisfy
\be
\label{VCT}
    V_{\rm CT} &= & - \frac{1}{2} \delta \mu_h^2 h^2 + \frac{1}{4} \delta \lambda_h h^4 + \delta m_{h \phi_1}^2 h \phi_1 + \delta m_{h \phi_2}^2 h \phi_2 \nonumber \\
    && + \delta m_{\phi_1 \phi_2}^2 \phi_1 \phi_2 + \frac{1}{4} \delta \lambda_{h \phi_1} h^2 \phi_1^2 + \frac{1}{4} \delta \lambda_{h \phi_2} h^2 \phi_2^2 \, .~~~~~~~
\ee
The various coefficients in this expression are fixed by imposing the following on-shell renormalization conditions at zero-temperature:
\begin{equation}
\label{Boundarycond}
\frac{\partial (V_\text{CW} + V_\text{CT})}{\partial y_{i}} =0\,,  \,\,\,\,\,\,\,\,\,
\frac{\partial^2 (V_\text{CW} + V_\text{CT})}{\partial y_{i} \partial y_{j}} =0\,,
\end{equation}
where $\{y_1, y_2, y_3 \}= \{\phi_1, \phi_2, h\}$. All first and second derivatives are evaluated at the true electroweak minimum field values, $\{ \phi_1, \phi_2, h\} = \{0, 0, v_h\}$. The coefficients of the counterterms are related to the derivatives of the potential as follows:
\be
\delta\mu_h^2  &=& \frac{3V_h}{2 v}  - \frac{V_{hh}}{2}\,, \,\,\,\,
\delta \lambda_{h} =\frac{V_h}{2 v^3}  - \frac{V_{hh}}{2 v_h^2}\, , \\
\delta m_{h \phi_1}^2 &=& - \frac{V_{\phi_1}}{v} \, ,
\delta m_{\phi_1 \phi_2}^2 = -  V_{\phi_1 \phi_2} \, , 
\delta m_{h \phi_2}^2 = - \frac{ V_{\phi_2}}{v} \, , ~~~~~~
\\
\delta \lambda_{h \phi_1} &=& - \frac{2V_{\phi_1 \phi_1}}{v_h^2} \,, \,\,\,\,
\delta \lambda_{h \phi_2} = - \frac{2V_{\phi_2 \phi_2}}{v_h^2}  \, ,
\ee
where we have defined
\be
V_{y_i} \equiv \frac{\partial V_\text{CW}}{\partial y_i}\, , \,\,\,\, V_{y_i y_j} \equiv \frac{\partial^2 V_\text{CW}}{\partial y_i  \partial y_j} \, .
\ee

Note that the Goldstone modes have vanishing masses at the true, zero-temperature electroweak minimum.
%as we opt the Landau gauge for this work.
This leads to an infrared divergence~\cite{Martin:2014bca,Elias-Miro:2014pca} stemming from the second derivative used in the renormalization conditions found in Eq.~\ref{Boundarycond}. To address this, an infrared regulator can be introduced by modifying the Goldstone mode masses as $m_G^2 \to m_G^2 + \mu_{\rm IR}^2$.
In this study, we follow the approach outlined in Refs.~\cite{Baum:2020vfl, Chatterjee:2022pxf,Ghosh:2022fzp, Roy:2022gop, Bittar:2025lcr, Roy:2025zvo} and set $\mu_{\rm IR}^2 = 1\,{\rm GeV}^2$.

At finite temperatures, the effective potential receives additional corrections which, at the one-loop level, are given by~\cite{Dolan:1973qd, Weinberg:1974hy, Kirzhnits:1976ts}
\be
\label{Vthermal}
    V_{\rm TH}(m_i^2, T) = \frac{T^4}{2\pi^2} \sum_i n_i J_{B,F}\left[ \frac{m_i^2(h, \phi_1, \phi_2)}{T^2} \right] . \, \, \, \, \, \, \, \, \, \, \, \, \, \, \,   
\ee
The thermal functions, $J_{B(F)}$, 
are defined as
\be
J_{B (F)}(y^2)=
\pm {\rm Re}
\int_0^{\infty}
		x^2 \ln
		\left(
			1 \mp \exp^{-\sqrt{x^2 + y^2}}
		\right)
d{x} \, ,~~~~~~~
\label{eq:jbjf}
\ee
where the lower (upper) sign corresponds to fermions (bosons),
and the sum is performed over all particles, as described in Eq.~\ref{VCW}.

In the high-temperature limit, $m_i^{2} \ll T^2$, the thermal functions can be expanded to $\mathcal{O}\left(y^6\right)$ as
\be
    \label{eq:JBapprox}
    J_{B}(y^2)  &\approx& -\frac{\pi^4}{45}+\frac{\pi^2}{12}y^2-\frac{\pi}{6}y^3-\frac{1}{32}y^4 \log\left(\frac{y^2}{a_B}\right)  , ~~~~~~~~\\
    \label{eq:JFapprox}
    J_{F}(y^2) &\approx& \frac{7\pi^4}{360}-\frac{\pi^2}{24}y^2-\frac{1}{32}y^4 \log\left(\frac{y^2}{a_F}\right) ,
\ee
where we have defined
\be
a_B= 16\pi^2 \exp\left(\frac{3}{2} - 2 \gamma_E\right)\, , \,\,\,\, \,\,
a_F=  \frac{a_B}{16}\, ,
\ee
and $\gamma_E \approx 0.577$ is the Euler-Mascheroni constant. 
The presence of the $-\pi y^3/6$ term in the high-temperature expansion of $J_B$ can result in an energy barrier between two degenerate vacua, thus allowing for a strong first-order phase transition. Note that such a cubic term appears only for bosonic degrees-of-freedom, as it arises from the (Matsubara) zero mode propagator. This term is associated with divergences in the infrared limit.
Thus, at high temperatures, the perturbative expansion of the effective potential breaks down. This issue can be resolved by resumming a set of higher-loop diagrams, commonly known as daisy contributions.

Various resummation techniques have been developed to address this infrared problem by incorporating higher-order thermal corrections~\cite{Gross:1980br, Parwani:1991gq, Arnold:1992rz, Boyd:1993tz} (for a more detailed discussion, see Ref.~\cite{Bittar:2025lcr}).
Among these, we adopt the Parwani prescription~\cite{Parwani:1991gq} in this work. We denote the temperature- and field-dependent masses as $M_i^2$, where
\begin{equation}
M_i^2 = ~ \text{Eigenvalues}[M_{kl}^2] \, , \,\,\,\, M_{kl}^2 \equiv m_{kl}^2 + \Pi(T^2)\, ,
\end{equation}
and $\Pi(T^2) = c_{kl} T^2$, where $c_{kl}$'s are the so-called daisy coefficients.
In this scheme, $m_i^2$ is replaced by the temperature-dependent mass, $M_i^2$, in the one-loop CW correction and in the one-loop thermal correction potential.
The temperature-dependent
effective potential at one-loop order is thus given by
\be
\label{VtotT}
V_\text{eff}= V_0 + V_{\rm CW} (M_i^2) + V_{\rm CT} + V_{\rm TH}(M_i^2, T) \, . ~~~~
\ee

The EWPT can be studied by monitoring the evolution of the minimum of this potential as a function of temperature.
The production of gravitational waves from a first-order phase transition will be discussed in Appendix~\ref{GWs_section}.

The field-dependent tree-level mass-squared matrix for the CP-even scalar sector can be obtained from the scalar potential:
\begin{align}
   m_{y_i y_j}^2 \equiv \frac{\partial^2 V_0}{\partial y_i \, \partial y_j}  \, ,
\end{align}
where $\{y_1, y_2, y_3\} = \{ \phi_1, \phi_2, h \}$. 
Its components in this basis are given explicitly by
\be
\label{cpevenmasssq}
m_{{h h}}^2 &=& - \mu_h^2 +  3 \lambda_h h^2  + \frac{\eta_I}{2}  \phi_1 \phi_2 + \frac{\kappa}{2} (\phi_1^2 + \phi_2^2) \nonumber \\ 
&&~~~
 + \frac{\eta_R}{2} ( \phi_1^2 - \phi_2^2)\, , \nonumber  \\
m_{{\phi_1 \phi_1}}^2 &=& m_{0}^2 +  \frac{\kappa + \eta_R}{2}  h^2 + \rho_{0_R}^2  + \frac{\lambda_{12}}{2}  \phi_2^2 + 3 \lambda_1 \phi_1^2  \, , \label{mH22-masssq} \nonumber \\
m_{\phi_2 \phi_2}^2 &=& m_{0}^2 +  \frac{\kappa - \eta_R}{2}  h^2 - \rho_{0_R}^2  + \frac{\lambda_{12}}{2}  \phi_1^2 + 3 \lambda_2 \phi_2^2  \, ,  \label{mH33-masssq} \nonumber \\
m_{{h \phi_1}}^2 &=&  m_{{\phi_1 h}}^2 = \kappa h \phi_1 + \eta_R h \phi_2 - \eta_I h \phi_2  \, , \quad\quad \nonumber \\
m_{{h \phi_2}}^2 &=&  m_{{\phi_2 h}}^2 = \kappa h \phi_2 - \eta_R h \phi_1 - \eta_I h \phi_1  \, , \quad\quad   \nonumber \\
m_{{\phi_1 \phi_2}}^2 &=&  m_{\phi_2 \phi_1}^2 = - \rho_{0{_I}}^2  - \frac{1}{2} \eta_I h^2 + \lambda_{12} \phi_1 \phi_2 \, . \quad\quad   \label{mH23-masssq}
\ee
The field-dependent squared masses of the goldstone modes, gauge bosons, and top quark are given by
\begin{align}
\label{gaugetopmass}
m^2_{G^0,G^{\pm}} &= - \mu_h^2 + \lambda_h h^2 \, , \,\,\,\, 
m^2_{W^{\pm}} = \frac{1}{4} g_2^2 h^2  \, , \,\,\,\,\,\, \nonumber \\  
m_{Z}^2 &= \frac{1}{4} (g_1^2 + g_2^2) h^2 \, , \,\,\,\,\,\,
m_{t}^2 =  \frac{1}{2} y_t^2 h^2  \, . \nonumber
\end{align}

The daisy coefficients can be derived using Eq.~\ref{eq:JBapprox} from the high-temperature expansion of the thermal one-loop potential:
\begin{equation}
c_{kl}  =   \left.\frac{1}{T^2}\frac{\partial^2   V_{\rm TH}}{\partial  y_k  \partial  y_l}\right|_{T^2  \gg m^2} \, ,
\end{equation}
where we have obtained 
%$\{y_1,y_2,y_3\} = \{h, \phi_1, \phi_2\}$, so we have  
%
\begin{align}
%\label{eq:daisy-coeff}
c_{{h h}}  &= \!\frac{1}{16} (3 g_2^2 + g_1^2)  +  \frac{1}{4} y_t^2  + \frac{1}{48}(24 \lambda_h \!+ \! 4  \kappa - \eta_I \!)  \, ,%\label{Eq:DaisyCoeffs_h} 
\nonumber \\
c_{\phi_1 \phi_1} &= \frac{1}{24} \left(8 \kappa + 4 \eta_R + 6 \lambda_1 + \lambda_{12} \right) \, ,%\label{Eq:DaisyCoeffs_phi1} 
\nonumber \\
c_{\phi_2 \phi_2} &= \frac{1}{24} \left(8 \kappa - 4 \eta_R + 6 \lambda_2 + \lambda_{12} \right)  \, .
\label{Eq:DaisyCoeffs}
\end{align}
 To obtain the eigenvalues of the temperature-dependent mass-squared matrix, these corrections, 
$c_{{kl}} T^2$, are added to the expressions for $m_{kl}^2$, as given in Eq.~\ref{cpevenmasssq}. Similarly, thermal correction, $c_{h h} T^2$, must be added to the mass-squared terms for the goldstone modes. 

The temperature-dependent squared mass of $W^{\pm}_L$ is given by
\begin{align}
M_{W^{\pm}_L}^2  =  {1 \over 4}  g_2^2  h^2  +  \frac{11}{6} g_2^2  T^2.
\end{align} 

Lastly, the thermal corrections for the longitudinal component of the $Z$ boson and photon 
fields can be obtained by 
diagonalizing the following matrix:
\beq
~~~~\frac{1}{4} h^2
\begin{pmatrix}
g_2^2 &  -g_2 g_1 \\
-g_1 g_2 & g_1^2
\end{pmatrix}
+
\begin{pmatrix}
\frac{11}{6} g_2^2 T^2 & 0 \\
0 & \frac{11}{6} g_1^2 T^2 
\end{pmatrix}\, .
\eeq
%
%%%%%%%%%%%%%%%%%%%%%%%%%%%%%%%%%%%%%%%%%%%%%%%%%%%%%%%%%%%%%%%
\section{Gravitational Wave Production from a First-Order Phase Transitions}\label{GWs_section}
%%%%%%%%%%%%%%%%%%

A first-order phase transition in the early universe could potentially result in an observable stochastic background of gravitational waves. Unlike gravitational waves originating from binary systems, the stochastic background is unpolarized and isotropic, with an amplitude that follows a Gaussian distribution~\cite{Allen:1997ad}. Consequently, it can be characterized by a two-point correlation function that is proportional to the power spectral density, $\Omega_{\text{GW}}{h}^2$. The detection of a stochastic gravitational wave background would generally require the identification of a cross correlation across multiple detectors~\cite{Caprini:2015zlo, Cai:2017cbj, Caprini:2018mtu, Romano:2016dpx, Christensen:2018iqi}.

The dimensionless parameter, $\alpha$, can be written as~\cite{Espinosa:2010hh}
\beq
\label{alpha}
\alpha \equiv \frac{\Delta \rho}{\rho_{\text{rad}}} = \frac{1}{\rho_{\text{rad}}}  \left( \frac{T}{4} \frac{d\Delta V}{d T} - \Delta V \right) \, ,
\eeq
where $\Delta V$ is the difference between the potential energies at the false and true minima, and all quantities in this expression are evaluated at the completion of the phase transition.
In the absence of appreciable entropy injection during the phase transition, we typically expect the phase transition to end at a temperature that is approximately equal to the nucleation temperature.
The nucleation temperature can be found by solving
the following equation:
\beq
\label{nucleation}
\int_{T_n}^{\infty}\frac{dT}{T}\frac{\Gamma(T)}{H(T)^4} \approx 1 \, ,
\eeq
where $\Gamma(T)$ is the tunneling probability from the false vacuum to the true vacuum per unit time and volume~\cite{Turner:1992tz}. To compute this quantity, it is necessary to solve for the bounce solution of the so-called Euclidean action, $S_3(T)$~\cite{Linde:1981zj}. For this purpose, we have employed the publicly available package,~\cosmotransitions~\cite{Wainwright:2011kj}.

Eq.~\ref{nucleation} identifies the temperature at which the nucleation probability of a single bubble within a horizon volume becomes approximately equal to unity. This approximately translates to the criterion, $S_3(T)/T \approx 140$. Solving this equation allows one to determine $T_n$, which corresponds to the maximum temperature at which $S_3/ T \lesssim 140$~\cite{Apreda:2001us}.

The inverse of the duration of the phase transition, $\beta$, is given by
\beq
\label{beta}
\beta = -\frac{d S_3}{dt} \simeq 
\frac{1}{\Gamma}\frac{d \Gamma}{dt}
= H T \frac{d(S_3/T)}{dT},
\eeq
where the quantities in this expression are evalated at the end of the phase transition.

The energy that is released during the phase transition is shared between the plasma's kinetic energy, which causes a bulk motion of the fluid in the plasma resulting in gravitational waves, and the heating of the plasma. The parameter, $\kappa_v$, quantifies the fraction of the energy in latent heat that is converted into the bulk motion of the fluid. This can be estimated by~\cite{Espinosa:2010hh} 
\begin{equation}\label{kappav}
\kappa_v \approx \left[ \dfrac{\alpha}{0.73+0.083\sqrt{\alpha}+\alpha}\right]\,.
\end{equation}
Finally, $\kappa_{t}$ is the fraction of $\kappa_v$ that goes into the generation of MHD turbulence in the plasma. It is expected that $\kappa_t \approx (0.05 - 0.1) \, \kappa_v$~\cite{Hindmarsh:2015qta}, and we adopt a value of $\kappa_t/\kappa_v = 0.1$ in our calculations.

Stochastic gravitational waves arising from a cosmological first-order phase transition can be generated through three primary mechanisms: bubble wall collisions, long-lasting sound waves in the plasma, and MHD turbulence. In the case of bubble collisions, gravitational waves are produced by the stress-energy tensor associated with the bubble wall~\cite{Kosowsky:1992rz, Kosowsky:1991ua, Kosowsky:1992vn, Caprini:2019egz}.
For a phase transition occurring in a thermal plasma, however, the contribution from bubble collisions is negligible compared to the total gravitational wave energy density~\cite{Bodeker:2017cim}.
In contrast, the bulk motion of the plasma gives rise to velocity perturbations which lead to the generation of sound waves in a relativistic particle medium. These sound waves carry away the majority of the energy that is released during the phase transition~\cite{Hindmarsh:2013xza, Giblin:2013kea, Giblin:2014qia, Hindmarsh:2015qta,Caprini:2015zlo, Schmitz:2020syl}.
Additionally, the percolation of the plasma can induce MHD turbulence
%, particularly MHD turbulence, 
due to the ionized nature of the plasma, further contributing to the production of gravitational waves~\cite{Caprini:2006jb, Kahniashvili:2008pf, Kahniashvili:2008pe, Kahniashvili:2009mf, Caprini:2009yp, Kisslinger:2015hua}. 

By combining the contributions from sound waves and MHD turbulence, one can estimate the total gravitational wave intensity, $\Omega_{\text{GW}}h^2$.
The contribution of sound waves to the overall gravitational wave power spectrum can be described using the following fitting formula~\cite{Hindmarsh:2019phv}
\be
\label{soundgw}
\Omega_{\text{sw}}h^2 = 3 \times 10^{-6} \, \Upsilon_{\rm sw} v_{w} \left(\frac{H}{\beta}\right) 
\left(\frac{\kappa_{v} \alpha}{1+\alpha}\right)^2 G(f) \,,  ~~~
\ee
where $H$ is the Hubble rate evaluated at the temperature at the end of gravitational waves production (which we take to be equal to $T_n$),  
and we define
\be
G(f) \equiv (f/f_\text{sw})^3 
\left[\frac{7}{4 + 3 (f/f_\text{sw})^2}\right]^{7/2}.
\ee 
The present day peak frequency of the sound wave contribution is given by
\be
\label{fsw}
f_{\text{sw}}=2\times10^{-5}\hspace{1mm} \text{Hz} \, \left( \dfrac{1}{v_{w}}\right)\left(\dfrac{\beta}{H} \right) \left(\dfrac{T_n}{100 \hspace{1mm} \text{GeV}} \right) \,.~~~~~
\ee
The finite lifetime of the sound waves leads to a reduction in $\Omega_{\text{sw}}h^2$, represented by the suppression factor, $\Upsilon(\tau_{\text{sw}})$, appearing in Eq.~\ref{soundgw}~\cite{Guo:2020grp, Hindmarsh:2020hop}.
This suppression factor is given by
\beq
\Upsilon_{ \rm sw}= 1 - \frac{1}{\sqrt{1 + 2 \tau_{\text{sw}} H}} \,.
\label{eq:upsilon}
\eeq
The lifetime, $\tau_{\text{sw}}$, is evaluated at the timescale over which the turbulence develops, approximately given by
$\tau_{\text{sw}} \sim R/\bar{U}_f$,
where $R$ is the mean bubble separation and  $\bar{U}_f$ is
the root-mean-square (RMS) of the fluid velocity~\cite{Pen:2015qta,Hindmarsh:2017gnf}. $R$ is related to the duration of the phase transition,
$R = (8\pi)^{1/3} v_w /\beta$~\cite{Hindmarsh:2019phv, Guo:2020grp}. On the other hand, from hydrodynamic analyses, Refs.~\cite{Hindmarsh:2019phv,Weir:2017wfa} have shown that the RMS fluid velocity can be expressed as $\bar{U}_f = \sqrt{3 \kappa_v \alpha/4}$. 
In the limit of $\tau_{\text{sw}} \rightarrow \infty$, $\Upsilon_{\rm sw}$ approaches an asymptotic value of unity.

The contribution from MHD turbulence to the production of gravitational waves can be expressed as~\cite{Caprini:2015zlo}
\begin{align}
\label{GWturb}
\Omega_{t}h^2 &= 3 \times 10^{-4} \left(\frac{H}{\beta}\right) \left(\frac{\kappa_{t} \alpha}{1+\alpha}\right)^{3/2} Q(f) \,,
\end{align}
where we have defined
\be
Q(f) = \frac{(f / f_{t})^3}{\left(1 + f / f_{t}\right)
^{11/3} \left(1 + 8 \pi f / f_{\star}\right)} \,,
\ee
and where the present day peak frequency of the turbulence-induced gravitational wave spectrum, $f_t$, is defined as
\beq
\label{peakfreqturb}
f_{t}= 3 \times 10^{-2}~\text{mHz} \, \left( \frac{1}{v_w}\right)\left(\frac{\beta}{H}\right)\left(\frac{T}{100~\text{GeV}}\right) \,,
\eeq
and where
\begin{equation}\label{eq:h-star}
f_{*}=1.7 \times10^{-5}\hspace{1mm} \text{Hz} \,\left(\dfrac{T_n}{100 \hspace{1mm} \text{GeV}} \right) \,.
\end{equation}

\end{document}